\documentclass[prd,showpacs,nofootinbib,preprint,10 pt]{revtex4-1}
\usepackage{amsmath,graphicx,color,epsfig}
\usepackage{slashed}
\begin{document}
\begin{titlepage}
\begin{center}
\setlength {\baselineskip}{0.3in} 
{\bf\Large\boldmath Analysis of Angular Observables of $\Lambda_b \to \Lambda (\to p\pi)\mu^{+}\mu^{-}$ Decay in Standard and  $Z^{\prime}$ Models}\\[15mm]
\setlength {\baselineskip}{0.3in}
{\large Aqsa Nasrullah$^{1}$, M. Jamil Aslam$^{1, 2}$ and Saba Shafaq$^{3}$}\\[5mm]
~{\it $^{1}$Department of Physics, Quaid-i-Azam University, Islamabad 45320, Pakistan. \\
$^2$ Institute of High Energy Physics, P.O. Box 918(4),
Chinese Academy of Sciences, Beijing 100049, China.\\

$^3$Department of Physics, International Islamic University, Islamabad 45320, Pakistan.}\\[5mm]
%

%
%
{\bf Abstract}\\[5mm]
\end{center} 
\setlength{\baselineskip}{0.2in} 
In 2015, the LHCb collaboration has measured the differential branching ratio $\frac{d{\mathcal{B}}}{dq^2}$, the lepton- and hadron-side forward-backward asymmetries, denoted by $A^\ell_{FB}$ and $A^{\Lambda}_{FB}$, respectively in the range $15 < q^2(=s) < 20$ GeV$^2$ with 3 fb$^{-1}$ of data. Motivated by these measurements, we perform an analysis of $q^2$ dependent $\Lambda_b \to \Lambda (\to p \pi ) \mu^+\mu^-$ angular observables at large- and low-recoil in the SM and in a family non-universal $Z^{\prime}$ model. The exclusive $\Lambda_{b}\to \Lambda$ transition is governed by the form factors and in the present study we use the recently performed high-precision lattice QCD calculations that have well controlled uncertainties especially in $15 < s < 20$  GeV$^2$ bin.  Using the full four-folded angular distribution of $\Lambda_b \to \Lambda (\to p \pi ) \mu^+\mu^-$ decay, first of all we focus on the calculations of experimentally measured $\frac{d{\mathcal{B}}}{ds},\; A^\ell_{FB}$ and $A^{\Lambda}_{FB}$ in the SM and compare their numerical values with the measurements in appropriate bins of $s$. In case of the possible discrepancy between the SM prediction and measurements, we try to see if these can be accommodated though the extra neutral $Z^{\prime}$ boson. We find that in the dimuon momentum range $15 < s < 20$  GeV$^2$ the value of $\frac{d{\mathcal{B}}}{ds}$ and central value of  $A^\ell_{FB}$ in $Z^{\prime}$ model is compatible with the measured values. In addition, the fraction of longitudinal polarization of the dimuon $F_{L}$ is measured to be $0.61^{+0.11}_{-0.14}\pm 0.03$ in $15 < s < 20$  GeV$^2$ at the LHCb. We find that in this bin the value found in the  $Z^{\prime}$ model is close to the observed values. After comparing the results of these observables, we have proposed the other observables such as ${\alpha}_{i}$ and $\alpha^{(\prime)}_{i}$ with $i =\theta_{\ell},\; \theta_{\Lambda},\; \phi,\;L,\; U$ and coefficients of different foldings $\mathcal{P}_{1, \cdots, 9}$ in different bins of $s$ in the SM and $Z^{\prime}$ model.  We illustrate that the experimental observations of the $s$-dependent angular observables calculated here in several bins of $s$ can help to test the predictions of the SM and unravel
NP contributions arises due to $Z^{\prime}$ model in the $\Lambda_b \to \Lambda (\to p \pi ) \mu^+\mu^-$ decays.

\end{titlepage}

\section{Introduction}
Rare decays involving $b-$quark, such as $b \to (s\; , d)\gamma$, $b \to (s\;, d)\ell^{+}\ell^{-}$, are of immense interest since last couple of decades. This is because of the fact that these decays are induced by flavor-changing-neutral-current transitions (FCNC) involving the quantum number transitions $|\Delta  Q |=0$ and $|\Delta B| = 1$. In the Standard Model (SM), the FCNC transitions are not allowed at the tree level but occur at loop level because of Glashow-Iliopoulos-Maiani (GIM) mechanism \cite{GIM}. This make them sensitive to the masses of particles that run in the loop, e.g. $m_t$ and $m_W$ in the SM. As a consequence, these decays play a pivotal role in the determination of Cabibbo-Kobayashi-Masakawa (CKM) \cite{CKM} matrix elements in an indirect way. In different extensions of the SM, there is a possibility that the new particles can also run in the SM loop diagrams making these rare decays sensitive to the masses and couplings of the new particles. Hence, rare decays provide us a rich laboratory to test the predications of the SM and help us to establish possible New Physics (NP) indirectly \cite{Hurth, Blake}.

As long as the inclusive radiative and semi-leptonic decays are concerned, there are hardly any open issues that could lead us towards the evidence of NP. However, the experimental precision is also limited at present and it is expected that these bound get improved significantly at Belle II \cite{Ali}.   The situation for the exclusive semileptonic $B-$meson decays is different and it has a lot of open issues. Among them the most pertinent one is the lepton-flavor universality (LFU), i.e., the couplings of gauge bosons in the SM are same for different families of leptons. This important prediction of the SM can be testes by measuring the ratio of the decay widths of $B \to K^{(*)} \mu^{+}\mu^{-}$ and $B \to K^{(*)} e^{+}e^{-}$, defined as:
\begin{equation}
	\mathcal{R}_{K^{(*)}} = \frac{B \to K^{(*)} \mu^{+}\mu^{-}}{B \to K^{(*)} e^{+}e^{-}} \label{Ratio}
\end{equation}
in specific bins of the dilepton invariant mass squared that is written as $s \in [s_{\text{min}}\;, s_{\text{max}}]$ from here onwards. As this ratio involve the same $B \to K^{(*)}$ transition, the hadronic uncertainties arising from the form factors cancels out to a good approximation. Therefore, any possible deviations from the SM predictions, i.e., the value of ratio different from one will hint towards the NP. In 2014, the LHCb collaboration has observed more than $2 \sigma$ mismatch between the experimental observations and the SM predictions in different bins of the square of momentum transfer $s = q^2$ \cite{LHCb-14}. This hints towards the breakdown of LFU of the SM, i.e., the couplings of gauge bosons with $\mu$ and $e$ are not the same \cite{Isidori-16, Virto-17}. There are also some other areas where tensions between the SM predication and the experimental observations is found, such as the $\mathcal{P}_5$ anomaly ($3.5\sigma$ in one bin $s \in [4.30, 8.68]$ GeV$^2$ \cite{LHCb-13} that correspond to the certain coefficient in the angular distribution of the $B \to K^{*}(\to K\pi)\mu^{+}\mu^{-}$ decay \cite{LHCb-13, Matias-13-1, Matias-13-2}. This anomaly was again observed at $3 \sigma$ in the data with 3 fb$^{-1}$
luminosity in the two bins $s \in [4, 6]$ GeV$^2$ and $s \in [6, 8]$ GeV$^2$  \cite{ LHCb-16} and this is later confirmed by Belle in the bin $s \in [4, 8]$ GeV$^2$ \cite{Belle-16}. The fact that this anomaly was
accompanied by a $2.9\sigma$ tension in the second bin of another angular observable called $\mathcal{P}_{2}$ \cite{18}. In addition, there is small but noticeable difference found in the branching ratio of $B \to K^{*}\mu^{+}\mu^{-}$ \cite{LHCb-13-4, LHCb-14-1, HPQCD} and $B_s \to \phi \mu^{+}\mu^{-}$ ($2.0\sigma$ larger than the SM prediction both in low and large $\phi$ recoil) \cite{LHCb-13-5, LHCb-15, Meinel}. Making use of the available data and motivated by these tantalizing anomalies observed in these $B$ decays, in addition to explain them in different beyond the SM scenarios \cite{Ali-references} the global analyses have also been carried out \cite{18, 19, 20, 21, 22, 23, 24, 25, 26, 27}. Incorporating the factoraziable (absorbed in the form factors) and non-factorizable contributions, these global analyses favor the negative shift in the Wilson coefficient $C_9$ to explain most of the data. However, before we could claim that these are indications of NP, we have to get full control on the possible hadronic uncertainties arising due to form factors in the exclusive decays \cite{28, 29, 30, 31, 32, 33, 34}. In order to establish the hints of NP, on the experimental side we need to have an improved statistical data that is expected at the Belle II and the LHCb, whereas on theoretical side we can study some other decays that are governed at quark level by $b \to s \ell^{+}\ell^{-}\;  (\ell =  \mu\; , \tau)$ transitions. 

In the present study, we have considered the $\Lambda_b \to \Lambda (\to p \pi) \ell^{+}\ell^{-}$ decay that is interesting to its own regard. On experimental side, this decay was first studied by CDF collaboration \cite{41} and later the LHCb has published the first measurement of differential branching ratio as well as the forward-backward asymmetry of final state muon i.e., the  $\mathcal{A}_{FB}$ \cite{43, 42}. Recently, the LHCb collaboration has made the observation of $CP$ violation and the asymmetries arising due to the angle between the $\mu^{+}\mu^{-}$ and $pK^{-}$ planes $(a_{\text{CP}}^{\hat{T}\text{odd}})$ in $\Lambda_{b}\to pK^{-}\mu^{+}\mu^{-}$ by analysing the data available at an integrated luminosity of $3$ fb$^{-1}$ \cite{LHCb-17CP}.  On theoretical front, at first in the decay $\Lambda_{b} \to  \Lambda \ell^{+}\ell^{-}$, the hadrons involved in the initial and final state are the baryons, therefore the study of such decays will help us to understand the helicity structure of the underlying effective Hamiltonian \cite{35, 36, 37}. Another added benefit is that the analysis of angular asymmetries in the sequential decay $\Lambda_b \to \Lambda(\to p\pi) \mu^{+}\mu^{-}$ is expected to complement the different angular asymmetries in the corresponding $B \to K^{*}(\to K\pi)\ell^{+}\ell^{-}$ decays \cite{38, 39, 40}. One important aspect is the stability of $\Lambda$ under strong interactions and the decay $\Lambda_b \rightarrow \Lambda (\rightarrow p \pi ) \ell^+ \ell^-$ is theoretically cleaner than the decay $ B \rightarrow K^* ( \rightarrow K \pi ) \ell^+ \ell^- $. Due to these facts, the decay $\Lambda_b \to \Lambda \ell^{+}\ell^{-}$ has been theoretically studied in a number of papers \cite{44, 45, 46, 47, 48, 49, 50, 51, 52, 53, 54, 55, 56, 57, 58, 59, 60, 61, 62, 63, 64, 65, 66, 67, 68, 69, 70, 71, 72, 73, 74, 75, 76, 77, 78, 79, 80, 81}.

Just like the exclusive decays of $B$-mesons, the decay $\Lambda_b \to \Lambda \ell^{+}\ell^{-}$ is prone by the uncertainties arising due to form factors. However, at present the $\Lambda_b \to \Lambda$ transition form factors are calculated using lattice QCD calculations with high precision \cite{82} and to have their profile in the full $q^2$ range, these form factors are extrapolated using the  Bourrely-Caprini-Lellouch parametrization \cite{83}. The lattice results are quite consistent
with the recent QCD light-cone sum rule calculation \cite{46} with an added benefit that is much smaller uncertainty in most of the kinematic range. However, in contrast to the $B$ decays, the QCD factorization is not fully developed for the $b$-baryon decays, therefore, we will not include these non-factorizable contributions in the present study. After having a control on the hadronic uncertainties that mimic in the form factors, the next choice is to find the observables that are relatively clean. In line with the $B \to K^{*}(\to K\pi)\mu^{+}\mu^{-}$ decays, we have calculated combinations of different angular observables in $\Lambda_b \to \Lambda(\to p\pi)\mu^{+}\mu^{-}$ decays, namely, forward-backward asymmetries  $(A^{\ell}_{FB}\;, A^{\Lambda}_{FB}\;, A^{\ell\Lambda}_{FB})$,  the longitudinal $(F_{L})$ and transverse $(F_{T})$ fractions of dimuon, the longitudinal asymmetry $\alpha_L$, the transverse asymmetry $\alpha_U$ and the observables named as $\mathcal{P}_{i}$'s that are derived from different foldings, in the SM at its first right.

It has been observed that in order to explain the $R_K$ anomaly in the $B \to K \ell^{+}\ell^{-}$ decays, the possible candidate is the $Z^{\prime}$ model \cite{Z-prime}. The economy of these $Z^{\prime}$ models is that they can be accommodate to the SM only by extending the electroweak SM group by an additional $U(1)^{\prime}$ gauge group to which the extra-gauge boson $Z^{\prime}$ is associated. Also, in the grand unification theories (GUTs) such as $SU(5)$ or string inspired $E_{6}$ models \cite{GS,GB,EN,JB,VB}, one of the relevant scenarios is the family non-universal $Z^{\prime}$ model \cite{EE,PL1} and the leptophobic $Z^{\prime}$ models \cite{JLL,BBL}. The direct signature of an extra $Z^{\prime}$ boson is still missing in the analysis of data taken so far at the LHC \cite{TGR} experiment, but we already had some indirect constraints on the couplings of $Z^{\prime}$ gauge boson through low energy processes that are crucial and
complementary for direct searches $Z^{\prime}\to e^{+}e^{-}$ at Tevatron \cite{AAb}. The additional interesting thing that the family non-universal $Z^{\prime}$ models have in their account is  the new CP-violating phase which has large effects on various FCNC processes \cite{PL1,PL2}, such as $B_{s}-\bar B_{s}$ mixing \cite{ConstrainedZPC1,PL3,UTfit,CB,DL} and
rare hadronic and $B$-meson decays \cite{VB1,VB2,IA}. 
As extending the SM group by an extra $U(1)^{\prime}$ gauge group does not change the operator basis of the SM, therefore, the $Z^{\prime}$ model belongs to a class of Minimal Flavor violating Models having its imprints in the Wilson coefficients that correspond to the SM operators. Keeping in view that among the different hadrons produced at the LHCb, almost $20\%$ will be the $\Lambda_b$ baryons, it is expected that in future the results of decay distributions and different angular asymmetries will be available with much better statistics. Therefore, in addition to the SM calculation of the different observables mentioned above, we have studied the impact of different $Z^{\prime}$ parameters on these observables in different bins of $s$. 

The paper is organized as follows: In section II,  theoretical framework for the decay $\Lambda_b \to \Lambda(\to p\pi)\ell^+ \ell^-$ has been discussed. Helicity amplitudes for the decay are written in terms of transition form factors and four fold differential decay distributions. After summarizing the Wilson coefficients and other parameters of the $Z^{\prime}$ model in section III, we have given the calculation of several observables that have been obtained using four-folded angular distributions in section IV. Section V presents the numerical analysis of the observables done in the SM and in the $Z^{\prime}$ model and here we compare the results of certain asymmetries with the measurements available from the LHCb experiment.  In addition to the tabular form of the results, these have also been plotted graphically in the same section. Finally, the main findings are concluded in the last section. 

\section{Effective Hamiltonian Formalism for the SM and $Z^\prime$ model}
The quark level decay governing the $ \Lambda_b\rightarrow \Lambda(\rightarrow p \pi)\mu^{+}\mu^{-}$ is $b\rightarrow s\mu^{+}\mu^{-}$. In this decay of $b-$baryon, the short distance effects are encoded in the Wilson coefficients, whereas the long distance contributions are incorporated through the four quark operators. After integrating out the heavy degrees of freedoms, $W^{\pm} $, $Z$ bosons and top quark, the SM effective Hamiltonian for these decays is
\begin{equation}
H^{eff}_{SM}= -\frac{4G_{F}}{ \sqrt{2}}%
\frac{\alpha _{e}}{4\pi} V_{tb}V_{ts}^{\ast }{\sum_{{i=7,9,10}} C_{i}\left( \mu \right)\;O_{i}}\label{effective-Hamiltonian}
\end{equation}%
where $G_{F}$ is the Fermi coupling constant, $V_{tb}V^{\ast}_{ts}$ are the CKM matrix elements, $\alpha_e$ is the fine structure constant, $C_{i}\left( \mu \right)$ with $i = 7,\;9,\;10$ are the Wilson coefficients corresponding to the electromagnetic operator $\mathcal{O}_7$ and semileptonic operators $ \mathcal{O}_{9,10}$ that are defined as:
\begin{eqnarray*}
	O_{7 }&=&\frac{m_{b}}{e}\left[ \overline{s}\sigma
	^{\mu \nu }P_{R }b\right] F_{\mu \nu },\text{ \ \ \ \ }O_{9}=%
	\left[ \overline{s}\gamma ^{\mu }P_{L}b\right] \left[ \overline{\ell}\gamma
	_{\mu }\ell\right] ,\text{ \ \ \ \ }O_{10}=\left[ \overline{s}\gamma ^{\mu
	}P_{L}b\right] \left[ \overline{\ell}\gamma _{\mu }\gamma _{5}\ell\right]. 
\end{eqnarray*}%
It has already been mentioned that QCD factorization at low $q^2$ is not fully developed for the hadronic $b$-baryon decay, therefore, we have ignored the non-factorizable contributions here. \footnote{In case of $B \to K^{*}\mu^{+}\mu^{-}$ decay, it is evident that the non-factorizable charm-loop effects (i.e., corrections that are not described using hadronic form factors) play a sizeable role in the low $q^2$ region \cite{34} and the same is expected in case of the decay under consideration. However, in the present study we shall neglect their contributions because there is no systematic framework available in which these non-factorizable charm-loop effects can be calculated in the baryonic decays \cite{46}. Therefore, our results at low $q^2$ are effected by the uncertainities due to these contributions. In the whole $q^2$ range, the effective Wilson coefficients are given in Eq. (\ref{WC}). According to  ref. \cite{82}, we use Eq. (\ref{WC}) in low and high $q^2$ region by increasing $5\%$ uncertainty. Thus having a control on the non-factorizable contributions in baryonic decays will help us to hunt the deviations from the SM predictions.} The factorizable non-local matrix elements of the four quark operators $\mathcal{O}_{1-6}$ and $\mathcal{O}_{8}^g$ are encoded into effective Wilson coefficients $C_7^{eff}(s)$ and $C_9^{eff}(s)$ where $s$ is dilepton squared mass $q^2$ $(q^{\mu} = p_1^{\mu} - p_2^{\mu})$. In high $q^2$ region, the Wilson coefficients $C_7^{eff}(s)$ and $C_9^{eff}(s)$  can be written as \cite{25}:
\begin{eqnarray}
	C_7^{eff}(s) &=&C_{7}-\frac{1}{3}\left(C_{3}+\frac{4}{3}C_{4}+20C_{5}+\frac{80}{3}C_{6}\right)-\frac{\alpha_s}{4\pi}\left[ (C_{1}-6C_{2})F^{(7)}_{1,c}(s)+C_{8}F^{(7)}_{8}(s)\right],\notag \\
	C_{9}^{eff} (s) & =& C_{9}+\frac{4}{3}\left(C_{3}+\frac{16}{3}C_{5}+\frac{16}{9}C_{6}\right) - h(0,s)\left(\frac{1}{2}C_{3}+\frac{2}{3}C_{4}+8C_{5}+\frac{32}{3}C_{6}\right)\notag \\
	&&-\left(\frac{7}{2}C_{3}+\frac{2}{3}C_{4}+38C_{5}+\frac{32}{3}C_{6}\right)h(m_{b},s)+\left(\frac{4}{3}C_{1}+C_{2}+6C_{3}+60C_{5}\right)h(m_{c},s)\notag \\
	&&-\frac{\alpha_{s}}{4\pi}\left[C_{1}F_{1,c}^{(9)}(s)+C_{2}F_{2,c}^{(9)}+C_{8}F_{8}^{(9)}(s)\right] \label{WC}
\end{eqnarray}
where $h(m_{q^{\prime}},s)$ with $q^{\prime} = b\;, c$ corresponds to the fermionic loop functions. These $h(m_{q^{\prime}},s)$ along with the functions $F_{8}^{(7,9)}$ and $F_{(1,2,c)}^{(7,9)}$ are calculated in refs. \cite{47, 84-1}.

Long ago Langacker and
Pl\"{u}macher have included a family non-universal $Z^{\prime }$ boson through additional $U(1)^{\prime }$ gauge symmetry \cite{PL1}.  In contrast to the SM, having non-diagonal chiral coupling matrix, in a family non-universal $Z^{\prime }$
model, the FCNC transitions $b\rightarrow s\ell ^{+}\ell ^{-}$ could be
induced at tree level. Ignoring the $Z-Z^{\prime }$ mixing along with the assumption that the couplings of right handed
quark flavors with $Z^{\prime }$ boson are diagonal and 
the effective Hamiltonian for  $b\rightarrow s\ell ^{+}\ell ^{-}$ transition corresponding to $Z^{\prime}$ boson becomes \cite{QCh,KCh1,VB3}:
\begin{eqnarray}
	\mathcal{H}^{Z^{\prime}}_{eff}(b\to
	s\ell^{+}\ell^{-})=-\frac{2G_{F}}{\sqrt{2}}V_{tb}V^{\ast}_{ts}\left[\frac{B_{sb}S^{L}_{\ell\ell}}{V_{tb}V^{\ast}_{ts}}(\bar{s}b)_{V-A}
	(\bar{\ell}\ell)_{V-A}+\frac{B_{sb}S^{R}_{\ell\ell}}{V_{tb}V^{\ast}_{ts}}(\bar{s}b)_{V-A}
	(\bar{\ell}\ell)_{V+A}\right].\label{zp}
\end{eqnarray}
In Eq. (\ref{zp}), $S_{\ell \ell }^{L}$ and $S_{\ell \ell }^{R}$ represent the couplings
of $Z^{\prime }$ boson with the left- and right-handed leptons, respectively. The corresponding off-diagonal left-handed coupling of quarks
with new $Z^{\prime }$ boson is taken care of $B_{sb}=|\mathcal{B}_{sb}|e^{-i\phi
	_{sb}}$ with $\phi_{sb}$ a new weak phase. In a more sophisticated form, the above Eq. (\ref{zp}) can be written as
\begin{eqnarray}
	\mathcal{H}^{Z^{\prime}}_{eff}(b\to
	s\ell^{+}\ell^{-})=-\frac{4G_{F}}{\sqrt{2}}V_{tb}V^{\ast}_{ts}\left[\Lambda_{sb}
	C_{9}^{Z^{\prime}}O_{9}+\Lambda_{sb} C_{10}^{Z^{\prime}}O_{10}\right],\label{2}
\end{eqnarray}
where
\begin{equation}
	\Lambda_{sb}=\frac{4\pi e^{-i\phi_{sb}}}{\alpha_{EM}V_{tb}V^{\ast}_{ts}}\; ;  \; \quad\quad\quad\quad C_{9}^{Z^{\prime}}=|\mathcal{B}_{sb}|S_{LL}\; ; \; \quad\quad\quad\quad C_{10}^{Z^{\prime}}=|\mathcal{B}_{sb}|D_{LL}\; , \label{C9C10}
\end{equation}
and
\begin{equation}
	S_{LL}=S^{L}_{\ell\ell}+S^{R}_{\ell\ell}\; ; \quad\quad\quad\quad\quad D_{LL}=S^{L}_{\ell\ell}-S^{R}_{\ell\ell}\; .\label{4}
\end{equation}
By comparing Eq. (\ref{effective-Hamiltonian}) and Eq. (\ref{2}) it can be noticed that except $C_{7}^{eff}$, which is absent in $Z^{\prime}$ model, the operator basis of the family non-universal $Z^{\prime}$ model is same as that of the SM for $O_{9\; , 10}$. Hence the contribution arising due to the extra $Z^{\prime}$ boson is absorbed in the Wilson coefficients $C^{eff}_{9}$ and $C_{10}$. 

The total amplitude for the decay $\Lambda_{b}\to\Lambda\ell^{+}\ell^{-}$  is the sum of the SM and $Z^{\prime}$ contribution and it can be formulated in terms of $\Lambda_{b}\to \Lambda$ matrix elements as
\begin{eqnarray}
	\mathcal{M}^{tot}(\Lambda_{b}\to\Lambda\ell^{+}\ell^{-})
	= -\frac{G_{F}\alpha}{2\sqrt{2}\pi}V_{tb}V_{ts}^{\ast}[\langle\Lambda(k)|\bar s\gamma_{\mu}(1-\gamma_{5})b|\Lambda_{b}(p)\rangle
	\{C_{9}^{tot}(\bar\ell\gamma^{\mu}\ell)+C_{10}^{tot}(\bar\ell\gamma^{\mu}\gamma^{5}\ell)\}\notag\\
	-\frac{2m_{b}}{q^{2}}C_{7}^{eff}\langle\Lambda(k)|\bar s i\sigma_{\mu\nu}q^{\nu}(1+\gamma^{5})b|\Lambda_{b}(p)\rangle\bar\ell
	\gamma^{\mu}\ell],\label{6}
\end{eqnarray}
where $C_{9}^{tot}=C_{9}^{eff}+\Lambda_{sb} C_{9}^{Z^{\prime}}$ and $C_{10}^{tot}=C_{10}^{SM}+\Lambda_{sb} C_{10}^{Z^{\prime}}$, with $C_{9}^{eff}$ defined in Eq. (\ref{WC}).

\section{Helicity amplitudes and Form Factors for $\Lambda_b \to \Lambda$ transitions}
The matrix elements for the $\Lambda_{b}\to \Lambda$ transition for different possible currents, can be straightforwardly parameterized in terms of the form factors. The helicity formalism provide a convenient way to describe these transformations. The helicity amplitudes $H^i(s_1,s_2)$ with $i$ corresponding to vector $(V)$, axial-vector $(A)$ tensor $(T)$ and axial-tensor $(T_5)$ currents can be written as \cite{38}:
\begin{eqnarray}
H^{V}\left( s_{1},s_{2}\right)  &\equiv &\epsilon _{\mu }^{\ast }(\lambda) \left\langle \Lambda \left( p_{2},s_{2}\right) |\overline{s}\gamma
^{\mu }b|\Lambda _{b}\left( p_{1},s_{1}\right) \right\rangle \notag \\
&=&f_{0}\left( s\right) \frac{m_{\Lambda _{b}}-m_{\Lambda }}{\sqrt{s}}\left[ 
\bar{u}\left( p_{2},s_{2}\right) u\left( p_{1},s_{1}\right) \right] 
+2f_{+}\left( s\right) \frac{m_{\Lambda _{b}}+m_{\Lambda }}{s_{+}}\left(
p_{2}\cdot \epsilon ^{\ast }\left( 0\right) \right) \left[ \bar{u}%
\left( p_{2},s_{2}\right) u\left( p_{1},s_{1}\right) \right]   \notag \\
&&+f_{\bot }\left( s\right) \left[ \bar{u}\left( p_{2},s_{2}\right)
\slashed{\epsilon} ^{\ast }\left( \pm \right) u\left( p_{1},s_{1}\right) \right] \label{HC-1}  \\
H^{A}\left( s_{1},s_{2}\right)  &\equiv &\epsilon _{\mu }^{\ast }(\lambda) \left\langle \Lambda \left( p_{2},s_{2}\right) |\bar{s}\gamma
^{\mu }\gamma _{5}b|\Lambda _{b}\left( p_{1},s_{1}\right) \right\rangle  
\notag \\
&=&-g_{0}\left( s\right) \frac{m_{\Lambda _{b}}+m_{\Lambda }}{\sqrt{s}}\left[
\bar{u}\left( p_{2},s_{2}\right) \gamma _{5}u\left( p_{1},s_{1}\right) %
\right] -2g_{+}\left( s\right) \frac{m_{\Lambda _{b}}-m_{\Lambda }}{s_{-}}\left(
p_{2}\cdot \epsilon ^{\ast }\left( 0\right) \right) \left[ \bar{u}%
\left( p_{2},s_{2}\right) \gamma _{5}u\left( p_{1},s_{1}\right) \right] 
\notag \\
&&+ g_{\bot }\left( s\right) \left[ \bar{u}\left( p_{2},s_{2}\right)
\slashed{\epsilon} ^{\ast }\left( \pm \right) \gamma _{5}u\left( p_{1},s_{1}\right) %
\right] \label{HC-2}  \\
H^{T}\left( s_{1},s_{2}\right)  &\equiv &\epsilon _{\mu }^{\ast }(\lambda) \left\langle \Lambda \left( p_{2},s_{2}\right) |\bar{s}i\sigma
^{\mu \nu }q_{\nu }b|\Lambda _{b}\left( p_{1},s_{1}\right) \right\rangle  
\notag \\
&=&-2h_{+}\left( s\right) \frac{s}{s_{+}}\left( p_{2}\cdot \epsilon ^{\ast
}\left( 0\right) \right) \left[ \bar{u}\left( p_{2},s_{2}\right)
u\left( p_{1},s_{1}\right) \right] - h_{\bot }\left( s\right) \left( m_{\Lambda _{b}}+m_{\Lambda }\right) 
\left[ \bar{u}\left( p_{2},s_{2}\right) \slashed{\epsilon} ^{\ast }\left( \pm
\right) u\left( p_{1},s_{1}\right) \right] \label{HC-3} \\
H^{T_5}\left( s_{1},s_{2}\right)  &\equiv &\epsilon _{\mu }^{\ast }(\lambda)
\left\langle \Lambda \left( p_{2},s_{2}\right) |\bar{s}i\sigma ^{\mu
	\nu }q_{\nu }\gamma _{5}b|\Lambda _{b}\left( p_{1},s_{1}\right)
\right\rangle   \notag \\
&=&-2\widetilde{h}_{+}\left( s\right) \frac{s}{s_{-}}\left( p_{2}\cdot
\epsilon ^{\ast }\left( 0\right) \right) \left[ \bar{u}\left(
p_{2},s_{2}\right) \gamma _{5}u\left( p_{1},s_{1}\right) \right] 
+\widetilde{h}_{\bot }\left( s\right) \left( m_{\Lambda _{b}}-m_{\Lambda
}\right) \left[ \bar{u}\left( p_{2},s_{2}\right) \slashed{\epsilon} ^{\ast
}\left( \pm \right) \gamma _{5}u\left( p_{1},s_{1}\right) \right] \notag \\
\label{HC-4}.
\end{eqnarray}%
where $p_{1}(s_1)$ and $p_{2}(s_2)$ are the momentum (spin) of $\Lambda_b$ and $\Lambda$, respectively. The dilepton polarization vector is written as $\epsilon_{\mu}^{\ast}(\lambda)$ with $\lambda = t\;, 0\;, \pm$ and their explicit definitions are given in ref. \cite{38} and are summarised in Appendix. 

 In Eqs. (\ref{HC-1}, \ref{HC-2}, \ref{HC-3}) and Eq. (\ref{HC-4}), the functions $f_{i}(s),\; g_{i}(s),\; h_{i}(s)$ and $\widetilde{h}_{i}(s)$ with $i= 0\;, +\;, \bot$ are the transition form factors. In the heavy quark spin symmetry, the symmetry where the spin of spectator-diquark remains same in initial and final state,  the number of form factors are reduced. The tensor form factors can be written in terms of vector and axial-vector form factors and with this symmetry we can also equate the longitudinal and transverse form factors. Thus it reduces the number of independent form factors to two i.e., the Isuger-wise relations $\xi_1$ and $\xi_2$. The form factors, being the non-perturbative quantities needed to be calculated in some model. In the decay under consideration here, we will use the form factors that are calculated in Lattice QCD with much better control on the various uncertainties. In full dilepton mass square range these can be expressed as \cite{82}:
\begin{equation}
f(s)= \frac{a^f_0 +a^f_1 z(s)+a^f_2 (z(s))^2}{1-s/(m^f_{pole})^2},
\end{equation}
where, the inputs $a^f_0,a^f_1$ and $a^f_2$ are summarized in Tables \ref{TableI} and \ref{TableII}. The parameter $z$ is defined as \cite{82}
\begin{equation}
z(s)= \frac{\sqrt{t_{+}-s}- \sqrt{t_{+}-t_{0}}}{\sqrt{t_{+}-s}+\sqrt{t_{+}-t_{0}}} \label{z-parameter}
\end{equation}
with $t_0=( m_{\Lambda_b}-m_\Lambda)^2$ and $t_+ = (m_B+m_K)^2$.
\begin{table}[tb]
\caption{The values of form factors along with uncertainties calculated in the framework of lattice QCD with $(2+1)$ flavor dynamics for $\Lambda_b \rightarrow \Lambda$ transition \cite{82}.}\label{TableI}
\begin{tabular}{|c|c|c|c|c|c|}
\hline
Parameter & \ \ Input & Parameter & Input & Parameter & Input   \\ 
\hline
$a^{f_+}_0$ & $0.4229 \pm 0.0274$ & $a^{f_+}_1$ & $-1.3728 \pm 0.3068 $ & $a^{f_+}_2$ & $ 107972\pm 1.1506$   \\ 
$a^{f_0}_0$ & $ 0.3604\pm 0.0277 $ & $a^{f_0}_1$ & $-0.9284 \pm 0.3453$ & $a^{f_0}_2$ & $ 0.9861\pm 1.1988$  \\ 
$a^{f_{\perp}}_0$ & $ 0.5748\pm 0.0353 $ & $a^{f_{\perp}}_1$ & $ -1.4781\pm 0.4030 $ & $a^{f_{\perp}}_2$ & $ 1.2496\pm 1.6396$\\ 
$a^{g_+}_0$ &$ 0.3522\pm 0.0205$  & $a^{g_+}_1$ & $-1.2968 \pm 0.2732 $ & $a^{g_+}_2$ & $ 2.7106\pm 1.0665$  \\ 
$a^{g_0}_0$ & $0.4059 \pm 0.0267$ & $a^{g_0}_1$ & $ -1.1622\pm 0.2929 $ & $a^{g_0}_2$ & $ 1.1490\pm 1.0327$  \\ 
$a^{g_{\perp}}_0$ & $ 0.3522\pm 0.0205 $ & $a^{g_{\perp}}_1$ & $-1.3607 \pm 0.2949 $ & $a^{g_{\perp}}_2$ & $ 2.4621\pm 1.3711 $ \\ 
$a^{h_+}_0$ & $ 0.4753\pm 0.0423$ & $a^{h_+}_1$ & $-0.8840 \pm 0.3997$ & $a^{h_+}_2$ & $ -0.8190\pm 1.6760 $  \\
$a^{h_{\perp}}_0$ & $ 0.3745\pm0.0313 $ & $a^{h_{\perp}}_1$ & $-0.9439 \pm 0.2766 $ & $a^{h_{\perp}}_2$ & $ 1.1606\pm 1.0757 $\\
$a^{\tilde{h}_+}_0$ & $ 0.3256\pm 0.0248$ & $a^{\tilde{h}_+}_1$ & $-0.9603 \pm 0.2303$ & $a^{\tilde{h}_+}_2$ & $ 2.9780\pm 1.0041 $  \\
$a^{\tilde{h}_{\perp}}_0$ & $0.3256 \pm 0.0248$ & $a^{\tilde{h}_{\perp}}_1$ & $-0.9634 \pm 0.2268$ & $a^{\tilde{h}_{\perp}}_2$ & $ 2.4782\pm 0.9549 $ \\
\hline
\end{tabular}%
\end{table}

\begin{table}[tb]
\caption{Pole masses for different form factors \cite{82}.}\label{TableII}
\begin{tabular}{|c|c|c|}
\hline
$f$ \ \ \ &  \ \ \ \ $J^P$ \ \ \ \ & \ \ \ $m^f_{pole} $ \ \ \ \\ \hline
$f_0$ & $0^+$ & $5.711$ \\
$f_+,f_{\perp},h_+,h_{\perp}$ & $1^- $ & $5.416$ \\
$g_0$ & $0^-$ & $5.367$ \\
$g_+,g_{\perp},\tilde{h}_+,\tilde{h}_{\perp}$ & $1^+ $ & $5.750$ \\
\hline
\end{tabular}%
\end{table}
\section{Angular Distribution and Physical Observables}
The four-folded angular distribution of the four-body $\Lambda_{b} \to \Lambda(\to p\pi)\mu^{+}\mu^{-}$ decay, with an unpolarized $\Lambda_{b}$ can be written in terms of $K_{l,m}$ where $l$ and $m$ denotes the relative angular momentum and its third component for $p \pi$ and $ \mu^+ \mu^-$ systems, respectively as \cite{38}
\begin{eqnarray}
\frac{d^4\Gamma}{ds \ d\cos{\theta_{\Lambda}} \ d\cos {\theta}_{\ell} \ d\phi} &=& \frac{3}{8 \pi} \bigg[K_{1s^{\prime}s^{\prime}} \sin^2{\theta_{\ell}}+K_{1cc} \cos^2{\theta_{\ell}}+ K_{1c} \cos{\theta_{\ell}} + (K_{2s^{\prime}s^{\prime}} \sin^2{\theta_{\ell}}+K_{2cc} \cos^2{\theta_{\ell}}+ K_{2c} \cos{\theta_{\ell}}) \cos{\theta_{\Lambda}} \notag \\
&&+(K_{3s^{\prime}c} \sin{\theta_{\ell}} \cos{\theta_{\ell}}+K_{3s^{\prime}} \sin{\theta_{\ell}}) \sin{\theta_{\Lambda}} \sin{\phi} + (K_{4s^{\prime}c} \sin{\theta_{\ell}} \cos{\theta_{\ell}}+K_{4s^{\prime}} \sin{\theta_{\ell}}) \sin{\theta_{\Lambda}} \cos{\phi}  \bigg] \label{6}
\end{eqnarray}
In Eq. (\ref{6}), $\theta_{\ell}$ and $\theta_{\Lambda}$ are the helicity angles, $\phi$ is the azimuthal angle and $s$ is the dilepton mass square. The different kinematic relations are defined in ref. \cite{38}. The different angular coefficients correspond to the particular values of the $(l\;, m)$, e.g., the coefficients of $\cos^2{\theta_l}$, $\sin^2{\theta_l}$ and $\cos{\theta_{\ell}}$ correspond to $K_{0,\;0}$ whereas coefficients of  $\cos^2{\theta_{\ell}} \cos{\theta_{\Lambda}}$, $\sin^2{\theta_{\ell}} \cos{\theta_{\Lambda}}$ and $\cos{\theta_{\ell}} \cos{\theta_{\Lambda}}$ correspond to $K_{1,\;0}$ and the last four terms corresponds to $K_{1,\;1}$. These angular parameters $K_{i\; j}$, where $i=1, \cdots , 4$ and $j = s^{\prime}s^{\prime}, cc, c, s^{\prime}c, s^{\prime}$ are functions of the square of momentum transfer $s$. In terms of the transversity amplitudes, their explicit expressions are summarized in Appendix.

From the four-fold angular decay distribution, a number of physical observables  can be obtained after integrating on different parameters among $\theta_{\ell}$, $\theta_{\Lambda}$, $\phi$ and $s$.
\begin{figure}
\centerline{\includegraphics[scale=0.50]{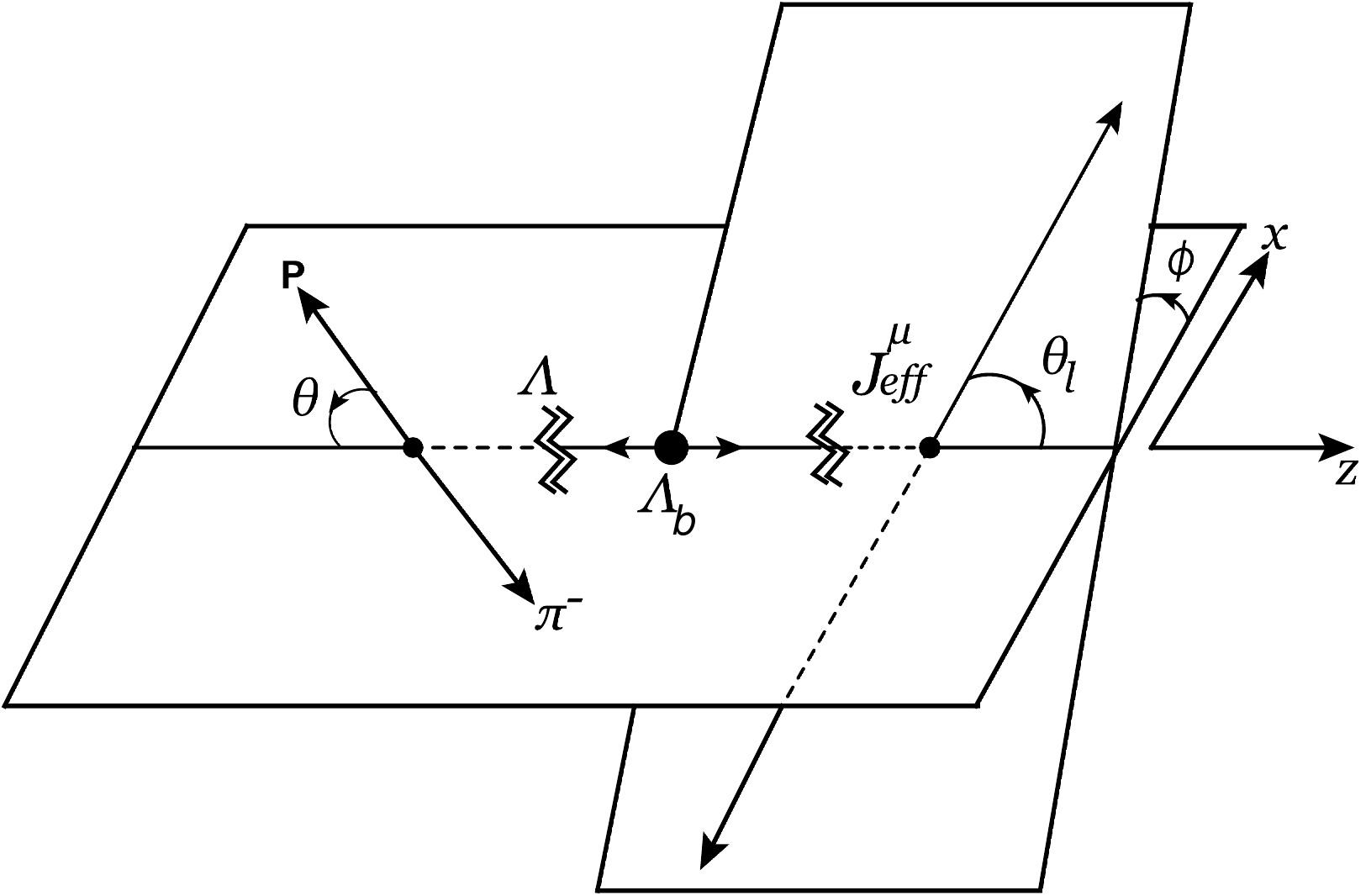}}
\caption{$\Lambda_{b} \to \Lambda(\to p\pi)\mu^{+}\mu^{-}$ decay topology, where $\theta_{\ell}$, $\theta = \theta_{\Lambda}$ are the helicity angles and $\phi$ is azimuthal angle.}
\label{geometry}
\end{figure}

\subsection{Differential decay rate and different asymmetry parameters}
One of the most important observables, both from the theoretical and experimental point of view, is the differential decay distribution. By integrating over $\theta_{\ell}\in [0, \; \pi]$, $\theta_{\Lambda}\in [0, \; \pi]$ and $\phi \in [0, \; 2\pi]$, the expression for the differential decay rate becomes
\begin{equation}
\frac{d\Gamma}{ds} = K_{1cc} + 2K_{1s^{\prime}s^{\prime}}\label{ddrate}
\end{equation}
In addition to the decay rate, we can extract a number of asymmetry parameters that correspond to different angels and they can be separated out by doing different integrations one by one. For example, by integrating on $\theta_{\ell}\in [0, \; \pi]$ and $\phi \in [0, \; 2\pi]$, the expression for the differential decay rate takes the form
\begin{equation}
\frac{d{\Gamma}}{ds d\cos{\theta_\Lambda}} = (K_{1cc} + 2K_{1s^{\prime}s^{\prime}})[1+\alpha_{\theta_{\Lambda}} \cos{\theta_{\Lambda}}]\label{drate-1}
\end{equation}
where $\alpha_{\theta_{\Lambda}}$ is the asymmetry parameter for the longitudinal polarization of the $\Lambda$ baryon. It can be noticed that if we integrate Eq. (\ref{drate-1}) on $\theta_{\Lambda}\in [0, \; \pi]$, we get back the Eq. (\ref{ddrate}). In terms of the helicity parameters $K_{i\;j}$, the asymmetry parameter  $\alpha_{\theta_{\Lambda}}$ can be expressed as follows:
\begin{eqnarray}
\alpha_{\theta_{\Lambda}}=\frac{\widetilde{K}_{2cc}+2 \widetilde{K}_{2s^{\prime}s^{\prime}}}{{K}_{1cc}+2 {K}_{1s^{\prime}s^{\prime}}}\label{asymmetry1}
\end{eqnarray}
 with $\widetilde{K}_{i,j} = \frac{K_{i,j}}{\alpha_{\Lambda}}$. Here $\alpha_{\Lambda}$ is the asymmetry parameter corresponding to the parity violating $\Lambda \to p \pi^{-}$ decay and its experimental value is $\alpha_{\Lambda} = 0.642\pm 0.013$ \cite{98}.
 
Similarly, by performing an integration on $\theta_{\Lambda} \in [0, \; \pi]$ and $\phi \in [0, \; 2\pi]$ and leaving the angle $\theta_{\ell}$, we will have asymmetries corresponding to angle $\theta_{\ell}$. In terms of $\alpha_{\theta_{\ell}}$ and $\alpha^{\prime}_{\theta_{\ell}}$, the differential decay rate can be formulated as
\begin{equation}
\frac{d{\Gamma}}{ds d\cos{\theta_{\ell}}} = K_{1s^{\prime}s^{\prime}} [1+\alpha_{\theta_{\ell}} \cos^{2}{\theta_{\ell}}+\alpha^{\prime}_{\theta_\ell} \cos{\theta_{\ell}}]\label{drate-2}
\end{equation}
with
\begin{eqnarray}
\alpha_{\theta_{\ell}}=\frac{K_{1cc}-K_{1s^{\prime}s^{\prime}}}{K_{1s^{\prime}s^{\prime}}},\quad\quad\quad\quad\quad
\alpha^{\prime}_{\theta_{\ell}}=\frac{{K}_{1c}}{K_{1s^{\prime}s^{\prime}}}\label{asymmetry2}
\end{eqnarray}
On the same lines, if we perform integration on the helicity angles $\theta_{\ell}\in [0, \; \pi]$ and $\theta_{\Lambda}\in [0, \; \pi]$, Eq. (\ref{6}) can be written in terms of asymmetries corresponding to the angle $\phi$ as
\begin{equation}
\frac{d{\Gamma}}{ds d{\phi}} = (K_{1cc} + 2K_{1s^{\prime}s^{\prime}})[1+\alpha_{\phi} \cos{\phi} + \alpha^{\prime}_{\phi} \sin{\phi}]\label{drate-3}
\end{equation}
where
\begin{eqnarray}
\alpha_{\phi} = \frac{3 \pi^2 \widetilde{K}_{4s^{\prime}}}{16 (K_{1cc}+2 K_{1s^{\prime}s^{\prime}})} ,\text{ \ \ \ \ \ \ \ \ \ \ \ }
\alpha^{\prime}_{\phi} = \frac{3 \pi^2 \widetilde{K}_{3s^{\prime}}}{16 (K_{1cc}+2 K_{1s^{\prime}s^{\prime}})}.
\end{eqnarray}
From Eq. (\ref{6}), the $s$ dependence of the transverse $(\alpha_{U})$ and longitudinal $(\alpha_{L})$ asymmetry parameters is written in the following form \cite{85}:
\begin{eqnarray}
\alpha_{U} = \frac{\widetilde{K}_{2cc}}{ K_{1cc}} ,\quad\quad\quad\quad\quad  \alpha_{L} = \frac{\widetilde{K}_{2s^{\prime}s^{\prime}}}{ K_{1s^{\prime}s^{\prime}}} \label{Long-Trans-asymmetry}
\end{eqnarray}

Even though one of the important observables is the decay rate, but it is a prone by the uncertainties arising from different input parameters where the major contributors are the form factors. It is a well established fact that the zero position of the forward-backward asymmetry in the different semileptonic decays of $B$-meson have a minimal dependence on the form factors \cite{Ali-FB}. Based on these observations the different forward-backward asymmetries are exploited in the $\Lambda_b$ decays \cite{38, 39, 40, 68, 69}. The forward-backward asymmetries corresponding to the lepton angle $\theta_{\ell}$ is defined as $A_{FB}^{\ell} = (F-B)/(F+B)$. Similarly, the hadron-side forward backward asymmetry, i.e., the asymmetry corresponding to the hardronic angle $\theta_{\Lambda}$ is $A_{FB}^{\Lambda} = (F-B)/(F+B)$. In both cases, $F$ and $B$ are the forward and backward hemispheres, respectively. From Eq. (\ref{6}), these forward-backward asymmetries become
\begin{equation}
A_{FB}^{\ell}=\frac{3K_{1c}}{4K_{1s^{\prime}s^{\prime}}+2K_{1cc}}   ,\text{
	\ \ \ \ }
A_{FB}^{\Lambda }=\frac{2K_{2s^{\prime}s^{\prime}}+K_{2cc}}{4K_{1s^{\prime}s^{\prime}}+2K_{1cc}} \label{AFB}
\end{equation}%
We take this opportunity to mention that in case of the $\Lambda_{b} \to \Lambda(\to p \pi)\mu^{+}\mu^{-}$ decay, the sequential decay $\Lambda \to p\pi$ is parity-violating. Therefore, the helicity components with the polarizations of proton to be $\pm \frac{1}{2}$ are not the same and hence the hadron-side forward-backward asymmetry is non-zero in these $b$-baryon decays. This is contrary to what we have seen in the $B \to K^{*}(\to K \pi)\mu^{+}\mu^{-}$ decay.
In addition to this, the combined lepton-hadron forward-backward asymmetry can be expressed as
\begin{equation}
A_{FB}^{\ell\Lambda }=\frac{3K_{2c}}{8K_{1s^{\prime}s^{\prime}}+4K_{1cc}} \label{combined-FB}
\end{equation}%

According to experimental point of view, the other interesting observables are the fractions of longitudinal $(F_{L})$ and transverse $(F_{T})$ polarized dimuons in $\Lambda_{b}\to \Lambda \mu^{+}\mu^{-}$ decay and these have already been measured in different bins by the LHCb Collaboration \cite{101}. In order to achieve the mathematical formula of these helicity fractions we have to integrate the four-folded differential decay rate given in Eq. (\ref{6}) on $\theta_{\Lambda} \in [0, \; \pi]$ and $\phi \in [0, \; 2\pi]$. Their explicit expressions in terms of $K_{i\;j}$ are
\begin{equation}
F_{T}=\frac{2K_{1cc}}{2K_{1s^{\prime}s^{\prime}}+K_{1cc}},\text{ \ \ \ \ }F_{L} =1-F_{T}=\frac{%
	2K_{1s^{\prime}s^{\prime}}-K_{1cc}}{2K_{1s^{\prime}s^{\prime}}+K_{1cc}}  
\end{equation}%
\subsection{Decay foldings and angular coefficients}
The four-fold decay distribution defined in Eq. (\ref{6}) gives us a chance to single out the different physical observables by studying different foldings. In semileptonic $B$-meson decays, such foldings have been studied in detail especially the penguin asymmetries $\mathcal{P}$, where $P^{(\prime)}_{5}$ is the most important \cite{18}. On the same lines, by using the foldings defined in Table \ref{Table-Foldings}, corresponding to different variations of azimuthal angle $\phi$ while taking $\theta_{\ell} \in [0, \frac{\pi}{2}]$ and $\theta_{\Lambda} \in [0 , \frac{\pi}{2}]$, these foldings can be expressed in terms of different angular coefficients as:
\begin{eqnarray}
\frac{d\Gamma_1}{\widehat{\Gamma}}&=& \frac{3}{8 \pi} \left[ 2 \frac{K_{1cc}}{ \widehat{\Gamma}}+ \mathcal{P}_1 \sin^2{\theta_{\ell}}+\frac{1}{2} \mathcal{P}_9 \cos{\theta_{\Lambda}} + \mathcal{P}_2 \sin^2{\theta_l} \cos{\theta_{\Lambda}} +\frac{1}{2} \mathcal{P}_8 \cos{\theta_{\ell}}+ \mathcal{P}_3 \cos{\theta_{\ell}} \cos{\theta_{\Lambda}} \right] \notag\\
\frac{d\Gamma_2}{\widehat{\Gamma}}&=& \frac{3}{8 \pi} \left[ 4 \frac{K_{1cc}}{ \widehat{\Gamma}}+2 \mathcal{P}_1 \sin^2{\theta_{\ell}}+ \mathcal{P}_6 \sin{\theta_{\ell}}  \sin{\theta_{\Lambda}} \cos{\phi}+ 2 \mathcal{P}_4 \sin{\theta_{\ell}} \cos{\theta_{\ell}} \sin{\theta_{\Lambda}} \sin{\phi} \right]\notag \\
\frac{d\Gamma_3}{\widehat{\Gamma}}&=& \frac{3}{8 \pi} \left[ 4\frac{K_{1cc}}{ \widehat{\Gamma}}+2 \mathcal{P}_1 \sin^2{\theta_{\ell}}+ \mathcal{P}_6 \sin{\theta_{\ell}}  \sin{\theta_{\Lambda}} \cos{\phi} + 2 \mathcal{P}_5 \sin{\theta_{\ell}} \cos{\theta_{\ell}} \sin{\theta_{\Lambda}} \cos{\phi} +  \mathcal{P}_8 \cos{\theta_{\ell}} \right]\notag \\
\frac{d\Gamma_4}{\widehat{\Gamma}}&=& \frac{3}{8 \pi} \left[ 4 \frac{K_{1cc}}{ \widehat{\Gamma}}+2 \mathcal{P}_1 \sin^2{\theta_{\ell}}+ \mathcal{P}_6 \sin{\theta_{\ell}}  \sin{\theta_{\Lambda}} \cos{\phi} +  \mathcal{P}_7 \sin{\theta_{\ell}}  \sin{\theta_{\Lambda}} \sin{\phi} \right]\notag \\
\frac{d\Gamma_5}{\widehat{\Gamma}}&=& \frac{3}{8 \pi} \left[ 4\frac{K_{1cc}}{ \widehat{\Gamma}}+2 \mathcal{P}_1 \sin^2{\theta_{\ell}}+ \mathcal{P}_6 \sin{\theta_{\ell}}  \sin{\theta_{\Lambda}} \cos{\phi} + 2 \mathcal{P}_3 \cos{\theta_{\ell}}  \cos{\theta_{\Lambda}} \right]\notag \\
\frac{d\Gamma_6}{\widehat{\Gamma}}&=& \frac{3}{8 \pi} \left[ 4 \frac{K_{1cc}}{ \widehat{\Gamma}}+2 \mathcal{P}_1 \sin^2{\theta_{\ell}}+ \mathcal{P}_9 \cos{\theta_{\Lambda}} + 2 \mathcal{P}_2 \sin^2{\theta_{\ell}} \cos{\theta_{\Lambda}} +  \mathcal{P}_6 \sin{\theta_{\ell}}  \sin{\theta_{\Lambda}} \cos{\phi} \right] \label{Folded-distributions}
\end{eqnarray}
Following things can be noticed from Eq. (\ref{Folded-distributions}):
\begin{itemize}
\item The coefficients of $\sin^2 \theta_{\ell}$ and $\sin^2 \theta_{\ell}\cos\theta_{\Lambda}$ correspond to the angular coefficients named as $\mathcal{P}_{1}$ and $\mathcal{P}_{2}$, respectively.
\item The coefficient of  $\cos \theta_{\ell}\cos\theta_{\Lambda}$ corresponds to the angular coefficient $\mathcal{P}_{3}$ and that of $\sin\theta_{\ell}\cos\theta_{\ell}\sin\theta_{\Lambda}\sin\phi$ is $\mathcal{P}_{4}$.
\item $\mathcal{P}_{5}$ is the coefficient of $\sin\theta_{\ell}\cos\theta_{\ell}\sin\theta_{\Lambda}\cos\phi$, where as $\mathcal{P}_{6}$ is the coefficient of $\sin\theta_{\ell}\sin\theta_{\Lambda}\cos\phi$.
\item $\mathcal{P}_{7}$, $\mathcal{P}_{8}$ and $\mathcal{P}_{9}$ are the coefficients of $\sin\theta_{\ell}\sin\theta_{\Lambda}\sin\phi$, $\cos\theta_{\ell}$ and $\cos\theta_{\Lambda}$, respectively.
\end{itemize}
In terms of the different helicity components, the angular coefficients $\mathcal{P}_{i}$, $i=1, . . . ,9$ are
\begin{eqnarray}
	\mathcal{P}_{1} &=&\frac{2}{ \widehat{\Gamma}}\left( K_{1ss}-K_{1cc}\right) ,%
	\text{ \ \ \ \ }\mathcal{P}_{2}=\frac{2}{ \widehat{\Gamma}}\left(
	K_{2ss}-K_{2cc}\right),  \text{ \ \ \ \ } 
	\mathcal{P}_{3} =\frac{2K_{2c}}{ \widehat{\Gamma}}, \text{ \ \ \ \ } \notag \\
	\mathcal{P}_{4}&=&\frac{2K_{3sc}}{ \widehat{\Gamma}},\text{ \  }
\hspace{2cm}	\mathcal{P}_{5}=\frac{2K_{4sc}}{ \widehat{\Gamma}}\text{, \ \ \ \ \ \ \ \ \ } 
\hspace{1cm}	\mathcal{P}_{6} =\frac{4K_{4s}}{ \widehat{\Gamma}} \notag \\
	\mathcal{P}_{7}&=&\frac{4K_{3s}}{ \widehat{\Gamma}}\text{, \ \ \ }  \hspace{2cm}  \mathcal{P}_{8}=\frac{4K_{1c}}{ \widehat{\Gamma}} \text{, \ \ \ \ \ \ \ \ \ \  }
	\hspace{1cm} \mathcal{P}_{9} =\frac{4K_{2cc}}{ \tilde{\Gamma}} \label{various-pasym}
\end{eqnarray}
where $\widehat{\Gamma}= \frac{d{\Gamma}}{ds} $.
It is worth mentioning that while obtaining the different $\mathcal{P}_{i}$'s, we have used the first six foldings defined in Table \ref{Table-Foldings}, because the last two foldings do not add any new observable. 

In order to compare the results with some of the experimentally measured observables and to propose possible candidates that might be useful to establish new physics, the interesting quantities are the normalized fractions calculated in different bins of square of dimuon momentum i.e., $s = q^2$. The normalized branching ratio, various asymmetry observables and different angular coefficients can be calculated as
\begin{equation}
\left\langle X \right\rangle  = \frac{\underset{s_{min}}{\overset{s_{max}}{\int}} X \ ds}{\underset{s_{min}}{\overset{s_{max}}{\int}} (\frac{d\Gamma}{ds})ds}.\label{normalized}
\end{equation}

\begin{table}
	\caption{Foldings required for $\mathcal{P}_i$'s for which $\theta_{\Lambda} \in [0, \frac{\pi}{2}]$, $\theta_l \in [0, \frac{\pi}{2}]$ and $\phi$ vary in different range corresponding to different observables \cite{18}.}\label{Table-Foldings}
\begin{tabular}{|l|l|l|}
\hline
Sr. no.&\ \ \ \ \ \ \ \ \ \ \ \ \ \ \ \ \ \ \ \ \ \ \ \ \ \ \ \ \ \ \ \ \ \ \ \ \
\ \ \ \ \ \ \ \ \ \ \ \ \ \ \ Folding & $\phi $ Range \\ \hline
1.& $d\Gamma \left( {\phi },{\theta }_{l},{\theta_{\Lambda} }%
\right) +d\Gamma \left( {\phi }-\pi ,{\theta }_{l},%
{\theta_{\Lambda}}\right) $ & $\left[ 0,\pi \right] $ \\ 
2.& $d\Gamma \left( {\phi },{\theta }_{l},{\theta_{\Lambda}}%
\right) +d\Gamma \left( {\phi },{\theta }_{l},\pi -%
{\theta }_{\Lambda}\right) +d\Gamma \left( -{\phi },\pi -{%
	\theta }_{l},{\theta }_{\Lambda}\right) +d\Gamma \left( -{\phi }%
,\pi -{\theta }_{l},\pi -{\theta }_{\Lambda}\right) $ & $\left[
0,\pi \right] $ \\ 
3.& $d\Gamma \left( {\phi },{\theta }_{l},{\theta }%
_{K}\right) +d\Gamma \left( {\phi },{\theta }_{l},\pi -%
{\theta }_{\Lambda}\right) +d\Gamma \left( -{\phi },{%
	\theta }_{l},{\theta }_{\Lambda}\right) +d\Gamma \left( -{\phi },%
{\theta }_{l},\pi -{\theta }_{\Lambda}\right) $ & $\left[ 0,\pi %
\right] $ \\ 
4.& $d\Gamma \left( {\phi },{\theta }_{l},{\theta }%
_{\Lambda}\right) +d\Gamma \left( {\phi },{\theta }_{l},\pi -%
{\theta }_{\Lambda}\right) +d\Gamma \left( {\phi },\pi -{%
	\theta }_{l},{\theta }_{\Lambda}\right) +d\Gamma \left( {\phi }%
,\pi -{\theta }_{l},\pi -{\theta }_{\Lambda}\right) $ & $\left[
0,\pi \right] $ \\ 
5.& $d\Gamma \left( {\phi },{\theta }_{l},{\theta }%
_{\Lambda}\right) +d\Gamma \left( -{\phi },{\theta }_{l},{%
	\theta }_{\Lambda}\right) +d\Gamma \left( {\phi },\pi -{\theta }%
_{l},\pi -{\theta }_{\Lambda}\right) +d\Gamma \left( -{\phi },\pi -%
{\theta }_{l},\pi -{\theta }_{\Lambda}\right) $ & $\left[ 0,\pi %
\right] $ \\ 
6.& $d\Gamma \left( {\phi },{\theta }_{l},{\theta }%
_{\Lambda}\right) +d\Gamma \left( -{\phi },{\theta }_{l},{%
	\theta }_{\Lambda}\right) +d\Gamma \left( {\phi },\pi -{\theta }%
_{l},{\theta }_{\Lambda}\right) +d\Gamma \left( -{\phi },\pi -%
{\theta }_{l},{\theta }_{\Lambda}\right) $ & $\left[ 0,\pi \right] 
$ \\ 
7.& $d\Gamma \left( {\phi },{\theta }_{l},{\theta }%
_{\Lambda}\right) +d\Gamma \left( \pi -{\phi },{\theta }_{l},%
{\theta }_{\Lambda}\right) +d\Gamma \left( {\phi },\pi -{%
	\theta }_{l},{\theta }_{\Lambda}\right) +d\Gamma \left( \pi -{\phi 
},\pi -{\theta }_{l},{\theta }_{\Lambda}\right) $ & $\left[ -\pi
/2,\pi /2\right] $ \\ 
8.& $d\Gamma \left( {\phi },{\theta }_{l},{\theta }%
_{\Lambda}\right) +d\Gamma \left( \pi -{\phi },{\theta }_{l},%
{\theta }_{\Lambda}\right) +d\Gamma \left( {\phi },\pi -{%
	\theta }_{l},\pi -{\theta }_{\Lambda}\right) +d\Gamma \left( \pi -{%
	\phi },\pi -{\theta }_{l},\pi -{\theta }_{\Lambda}\right) $ & $%
\left[ -\pi /2,\pi /2\right] $ \\ \hline
\end{tabular}
\end{table}

\section{Numerical Analysis}
In this section we discuss the numerical results obtained for different observables defined in Section-IV both in the SM and $Z^{\prime}$ model for the $\Lambda_{b}\to \Lambda(\to p \pi)\mu^{+}\mu^{-}$ decay. In $\Lambda_b \to \Lambda$ decays, the final state $\Lambda \to p \pi^{-}$ is a parity-violating decay and the corresponding asymmetry parameter $(\alpha_{\Lambda})$ is measured experimentally \cite{98}. This is really helpful in disentangling the direct $\Lambda_{b} \to p\pi^{-}\mu^{+}\mu^{-}$ from the one that is occurring through the intermediate $\Lambda$ decay that subsequently decays to $p\pi^{-}$. This is contrary to $B \to K^{*}(\to K\pi)\mu^{+}\mu^{-}$ decay where the final state $K^{*}$ meson decays to $K\pi$ via strong interaction. Therefore, the angular analysis of $\Lambda_{b}\to\Lambda(\to p\pi^{-})\mu^{+}\mu^{-}$ decay is quite interesting both theoretical and experimental point of views \cite{38, 39}. 
In addition to the above given input parameters, the other important ingredient in the numerical calculations in $\Lambda_{b}$ decays is the form factors. In the numerical calculation, we will use one of the most accurately calculated form factors at the QCD lattice \cite{82} with 2+1 flavor dynamics (c.f. Table \ref{TableI}) along with the NNLL corrections to the form factors for the SM that are given in \cite{84-1, 99}.
 
In addition to the form factors, the numerical values of the other input parameters that correspond to the SM and $Z^{\prime}$ model are given in Tables \ref{SM-values} and \ref{Zprime-values}, respectively. Using these values a quantitative analysis of above calculated observables in various bins of $s$ is presented in Tables \ref{low-high-bin-values}, \ref{Bin-analysis} and \ref{pbin-values}. In the whole analysis, we have observed that the results are not sensitive to the different scenarios of the $Z^{\prime}$ model, therefore, we have used only the scenario $\mathcal{S}_{1}$ to generate the results in various bins of $s$.

\begin{table}
		\caption{Numerical values of the different input parameters corresponding to the SM \cite{82, 98}. The Wilson coefficients at the scale $\mu_{b} = 4.2$ GeV, to NNLL accuracy in the SM \cite{Hu:2017qxj}.}\label{SM-values}
		\begin{tabular}{|c|c|c|c|c|c|}
	\hline
	 \ \ \ \ \ $G_F$ \ \ \ \ \  & $1.16638\times 10^{-5}$ & \ \ \ $\alpha_s(m_Z)$ \ \ \ &  \ \ \ \ 0.1182 \ \ \ & \ \ \ \ $m^{pole}_b$ \ \ \ \ & \ \ \ \ \ 4.78 \ \ \ \ \  \\ 
	 	 $\mu$ & 4.2 & $m_Z$ & 91.1876 & $\alpha_e$ & $\frac{1}{128}$   \\ 
	 $m_{\pi}$ & 0.135 & $m_K$ & 0.494  &$m_{\Lambda_b}$ & 5.619 \\ 
	 $m_{B}$ & 5.279 & $V_{tb}V^*_{ts}$ & 0.04152  & $m_{\Lambda}$ & 1.116   \\
	 $\tau_{\Lambda_b}$ & $1.466ps$ & $\alpha_{\Lambda}$ & $0.64 $ &$C_1$ & $-0.294$   \\ 
	  $C_2$ & 1.017 &$C_3$ & $-0.0059$ & $C_4$ & -0.087  \\ 
	 $C_5$ & 0.0004 & $C_6$ & 0.001 &$C_7$ & $-0.324$  \\ 
	 $C_8$ & $-0.176$& $C_9$ & $4.114$ & $C_{10}$ & $-4.193$  \\
	\hline
	\end{tabular}%
\end{table}

\begin{table}
	\caption{Numerical values of the parameters corresponding to the different scenarios of $Z^{\prime}$ model \cite{IA}.}\label{Zprime-values}
		\begin{tabular}{c|c|c|c|c}
		\hline
	    \ \ \ \ & \ \ \ \ $|\mathcal{B}_{sb}| \times 10^{-3}$ \ \ \ \ & \ \ \ \ $\phi_{sb} $ (in degrees) \ \ \ \ & \ \ \ \ $S_{LL} \times 10^{-2}$ \ \ \ \ & \ \ \ $ D_{LL} \times 10^{-2}$ \ \  \\ \hline
		$\mathcal{S}_1$ &$1.09\pm 0.22$ & $-72\pm 7$ & $-2.8\pm3.9$ & $-6.7\pm2.6$   \\ 
		$\mathcal{S}_2$ &$2.20\pm 0.15$ & $-82\pm 4$ & $-1.2\pm1.4$ & $-2.5\pm0.9$   \\ 
		$\mathcal{S}_3$ & $4.0\pm 1.5$ & $150\pm 10$ & 0.8 & $-2.6$   \\ 
		\hline
		\end{tabular}%
\end{table}

The first observable that is of the prime interest both from the theoretical and experimental point of views is the branching ratio in different bins of $s$ that can be set up by the experimentalists. From Eq. (\ref{ddrate}), in a bin $s \in [1.1, 6]$ GeV$^2$ (large-recoil) the average value of branching ratio in the SM and $Z^{\prime}$ model read
\begin{eqnarray}
\langle \mathcal{B}r \rangle_{\text{SM}} & = & (0.466^{+0.760}_{-0.394}) \times 10^{-7}, \nonumber \\
\langle \mathcal{B}r \rangle_{Z^{\prime}} & = & (0.709^{+0.115}_{-0.601}) \times 10^{-7}, \label{averagebr}
\end{eqnarray} 
whereas the measured value at the LHCb experiment in this particular bin is \cite{101}
\begin{equation}
\langle \mathcal{B}r \rangle_{exp}  =  (0.09 ^{+0.06}_{-0.05}) \times 10^{-7}.\label{expbr}
\end{equation}
By looking at Eqs. (\ref{averagebr}) and (\ref{expbr}), we can say that the deviations from the measured value in this bin are quite large in the SM and ever larger in the  $Z^{\prime}$ model. One of the possible reason of such large deviation is that the form factors are not very precisely calculated in this region. Contrary to this, the calculation of form factors is more precise in $s \in [15, 20]$ GeV$^2$ (low-recoil). In this bin the average value of  branching ratio in the SM and $Z^{\prime}$ model become
\begin{eqnarray}
\langle \mathcal{B}r \rangle_{\text{SM}} & = & (0.731 ^{+0.198}_{-0.187})\times 10^{-7},\nonumber \\
\langle \mathcal{B}r \rangle_{Z^{\prime}} & = & (1.179 ^{+0.271}_{-0.233}) \times 10^{-7}.  \label{averagebr-1}
\end{eqnarray} 
The experimentally measured value in this bin is \cite{101}
\begin{equation}
\langle \mathcal{B}r \rangle_{exp}  =  (1.20 \pm 0.25) \times 10^{-7}.\label{expbr-1}
\end{equation}
It can be reconciled that in this region, the deviations form the measured values are small compared to that of the large-recoil bin and in this case, the deviations are  $3.2 \sigma$ and $0.1 \sigma$ in the SM and $Z^{\prime}$ model, respectively. Hence, the results of branching ratio in $Z^{\prime}$ model for low-recoil bin look more promising when compared with the corresponding experimental value. In future, when we have more data from the LHCb experiment and Belle II, on one hand it will give us a chance to see the possible hints of the extra neutral $Z^{\prime}$ boson and on the other hand it help us to test the SM predictions with better accuracy.

It is well known fact that the branching ratio is prone by the uncertainties arising due to the form factors. In order to cope with some of the uncertainties, there are the observables such as the $\Lambda$ baryon forward-backward asymmetry $(A^{\Lambda}_{FB})$ and lepton forward-backward asymmetry $(A^{\ell}_{FB})$ that are measured w.r.t. the baryon angle $\theta_{\Lambda}$ and lepton angle $\theta_{l}$, respectively.  The asymmetry $A^{\Lambda}_{FB}$ can be expressed in terms of the ratio of  linear combination of the angular coefficients $K_{2ss}$ and $K_{2cc}$ to that of the linear combination of the angular coefficients $K_{1ss}$ and $K_{1cc}$ as given in Eq. (\ref{AFB}). Due to the change in the value of Wilson coefficient $C_9$ in $Z^{\prime}$ model, $K_{2ss}$ and $K_{2cc}$ get more contribution as compared to the $K_{1ss}$ and $K_{1cc}$. Hence, it will result in the significantly different values of $(A^{\Lambda}_{FB})$ in the SM and $Z^{\prime}$ model. In the first large-recoil bin $s\in[0.1,2]$ GeV$^2$, our results for $(A^{\Lambda}_{FB})$ in the SM and $Z^{\prime}$ model are
\begin{eqnarray*}
\langle A^{\Lambda}_{FB} \rangle_{\text{SM}}  =  -0.311 ^{+0.002}_{-0.001}\; , \text{ \ \ \ \ \ \ \ \ \ \ \ \ \ \ }  \langle A^{\Lambda}_{FB} \rangle_{Z^{\prime}}  =  -0.067 ^{+0.009}_{-0.002}\; ,
\end{eqnarray*} 
whereas the experimental result in this bin is \cite{101}
\begin{eqnarray*}
 \langle A^{\Lambda}_{FB} \rangle_{exp}=-0.12^{+0.31}_{-0.28}\; .
\end{eqnarray*}
It can be noticed that in comparison with the central values of experimental measurements in $s\in[0.1,2]$ GeV$^2$, the value of $Z^{\prime}$ model is $1.8$ times smaller, whereas, the one in the SM is $2.5$ times higher. In low-recoil region $(s\in[15, 20]$ GeV$^2$) the calculated values of $(A^{\Lambda}_{FB})$ are
\begin{eqnarray*}
\langle A^{\Lambda}_{FB} \rangle_{\text{SM}}  =  -0.273 ^{+0.003}_{-0.002}\; , \text{ \ \ \ \ \ \ \ \ \ \ \ \ \ \ }  \langle A^{\Lambda}_{FB} \rangle_{Z^{\prime}}  =  -0.137 ^{+0.001}_{-0.001}
\end{eqnarray*}
and the experimental value in this particular bin is \cite{101}
\begin{eqnarray*}
	\langle A^{\Lambda}_{FB} \rangle_{exp}=-0.29^{+0.07}_{-0.07}\; .
\end{eqnarray*}
It can be easily seen that at low-recoil, the SM prediction is close to experimentally measured value and the deviation is $0.2 \sigma$. The $Z^{\prime}$ value of $(A^{\Lambda}_{FB})$ exceeds from experimental result by $2.2 \sigma$. From the above discussion, it is clear that in the first bin of large-recoil both the SM and $Z^{\prime}$  model values deviate significantly from the experimental result of this bin,  whereas at low recoil the SM prediction is much closer to experimental result as compared to  $Z^{\prime}$ model. We hope in future, when more data will come from the LHCb, the results of measurements will become more concrete to compare with the SM and too see if the deviations can be accommodated with the $Z^{\prime}$ model. 

Another observable which is clean from the QCD uncertainties and that has been experimentally measured is the leptons forward-backward asymmetry $(A^{\ell}_{FB})$ which is an asymmetry w.r.t. the lepton scattering angle $(\theta_{l})$ and its mathematical expression is given in Eq. (\ref{AFB}). Here, it can be noticed that $(A^{\ell}_{FB})$ depends on the angular coefficient $K_{1c}$ and its denominator is same as that of $A^{\Lambda}_{FB}$. The angular coefficient $K_{1c}$  is higher for the SM than $Z^{\prime}$ model for $s< 4\;$GeV$^2$, whereas its behavior reverses when $s>4\;$ GeV$^2$. For $s<4\;$ GeV$^2$,  $K_{1c}$ is dominated by $C^{Z^{\prime}}_9$ whereas for $s>4\;$ GeV$^2$, the terms containing $C^{Z^{\prime}}_7$ dominate over the one that contain $C^{Z^{\prime}}_9$. Therefore, $A^{\ell}_{FB}$ increases with $s$ at the start of large-recoil and then it starts decreasing and crosses zero point at around $4\; $GeV$^2$. Our results for $\langle A^{\ell}_{FB}\rangle$ in the SM and $Z^{\prime}$ model calculated in an experimentally set-up bin $[0.1,2]$ GeV$^2$ are
\begin{eqnarray*}
\langle A^{\ell}_{FB} \rangle_{\text{SM}}  =  0.083 ^{+0.001}_{-0.035}\; , \text{ \ \ \ \ \ \ \ \ \ \ \ \ \ \ }  \langle A^{\ell}_{FB} \rangle_{Z^{\prime}}  =  0.040 ^{+0.003}_{-0.024}\; .
\end{eqnarray*}
The experimental value of $(A^{\ell}_{FB})$ in the corresponding bin is \cite{101}
\begin{eqnarray}
\langle A^{\ell}_{FB} \rangle_{\text{exp}}  =  0.37^{+0.37}_{-0.48} \label{AFBLvalue}.
\end{eqnarray}
In Eq. (\ref{AFBLvalue}), one can see that the errors are significantly large and it is likely to have an improvement from the future data of LHCb. However, the current central values are significantly away from the SM and $Z^{\prime}$ values, respectively. In low-recoil region $(s\in[15,20]$ GeV$^2)$ this asymmetry results
\begin{eqnarray*}
	\langle A^{\ell}_{FB} \rangle_{\text{SM}}  =  -0.180 ^{+0.007}_{-0.005}\; , \text{ \ \ \ \ \ \ \ \ \ \ \ \ \ \ }  \langle A^{\ell}_{FB} \rangle_{Z^{\prime}}  =  -0.135 ^{+0.003}_{-0.002}
\end{eqnarray*}
and to compare with the corresponding experimental value in this bin is
\begin{eqnarray*}
\langle A^{\ell}_{FB} \rangle_{\text{exp}}  =  -0.05^{+0.09}_{-0.09}.
\end{eqnarray*} 
It can be extracted that in this particular bin the average value of $ A^{\ell}_{FB}$ in $Z^{\prime}$ is comparable to the lower limit of the experimentally measured value, i.e., $-0.14$. 

In the category of the forward-backward asymmetry, the last is the combined forward-backward asymmetry $A^{\ell\Lambda}_{FB}$ which mainly depends on the angular coefficient $K_{2c}$ (c.f. Eq. (\ref{combined-FB})). Compared to the SM, the value of $K_{2c}$ is higher in the $Z^{\prime}$ model. 
At large-recoil our results in the SM and $Z^{\prime}$ model are
\begin{eqnarray*}
	\langle A^{\ell\Lambda}_{FB} \rangle_{\text{SM}}  =  -0.011 ^{+0.003}_{-0.006}\; , \text{ \ \ \ \ \ \ \ \ \ \ \ \ \ \ }  \langle A^{\ell\Lambda}_{FB} \rangle_{Z^{\prime}}  =  -0.009 ^{+0.002}_{-0.003}\; ,
\end{eqnarray*}
whereas at the low-recoil, the combined hadron-lepton forward-backward asymmetry is
\begin{eqnarray*}
	\langle A^{\ell\Lambda}_{FB} \rangle_{\text{SM}}  =  0.069 ^{+0.002}_{-0.002}\; , \text{ \ \ \ \ \ \ \ \ \ \ \ \ \ \ }  \langle A^{\ell\Lambda}_{FB} \rangle_{Z^{\prime}}  =  0.087 ^{+0.001}_{-0.002}\; .
\end{eqnarray*}
It can be seen that at large-recoil the deviations between the SM and $Z^{\prime}$ model is small and it grows significantly in the low-recoil region.

The next observable to be discussed here is the fraction of longitudinal polarization $(F_L)$ of the dilepton system. Due to linear combinations of same angular coefficients ($K_{1ss}$ and $K_{1cc}$) in both numerator and denominator of $F_L$, $Z^{\prime}$ model doesn't make much difference from the SM. The values in one of the large-recoil bin $[0.1,2]$ for the SM and $Z^{\prime}$ model are
\begin{eqnarray*}
	\langle F_{L} \rangle_{\text{SM}}  =  0.576 ^{+0.031}_{-0.174}\;, \text{ \ \ \ \ \ \ \ \ \ \ \ \ \ \ }  \langle F_{L} \rangle_{Z^{\prime}}  =  0.463 ^{+0.018}_{-0.095}
\end{eqnarray*}
and the corresponding experimental result is
\begin{eqnarray*}
	\langle F_{L} \rangle_{\text{exp}}  =  0.56 ^{+0.23}_{-0.56} \;.
\end{eqnarray*}
It can be observed that it is  in good agreement with the SM value and somewhat differ from the corresponding value in  $Z^{\prime}$ model for this bin. 

At low-recoil, the values of SM and $Z^{\prime}$ are
\begin{eqnarray*}
	\langle F_{L} \rangle_{\text{SM}}  =  0.713 ^{+0.010}_{-0.008}\;, \text{ \ \ \ \ \ \ \ \ \ \ \ \ \ \ }  \langle F_{L} \rangle_{Z^{\prime}}  =  0.590 ^{+0.007}_{-0.005}
\end{eqnarray*}
and the corresponding experimental result for this bin is
\begin{eqnarray*}
	\langle F_{L} \rangle_{\text{exp}}  =  0.61 ^{+0.11}_{-0.14}\;. 
\end{eqnarray*}
In contrast to the large-recoil, at low-recoil the results of $F_{L}$ in $Z^{\prime}$ model are closer to experimentally measured results. Therefore, to uncover the imprints of neutral boson in the longitudinal helicity fraction of dimuon system in $\Lambda_{b} \to \Lambda (\to p\pi) \mu^{+}\mu^{-}$ decays, the low-recoil bin might provide us a fertile ground.

Having a comparison of the SM and $Z^{\prime}$ model with the experimentally measured values of the different observables discussed above, we will now exploit some other observables that may be of interest in future at the LHCb and different B-factories. In connection with the $F_{L}$, the fraction of transverse polarization $F_T$ depends on $K_{1cc}$ and $K_{1ss}$ and its value at the large recoil is 
\begin{eqnarray*}
	\langle F_{T} \rangle_{\text{SM}}  =  0.136 ^{+0.021}_{-0.002}\;, \text{ \ \ \ \ \ \ \ \ \ \ \ \ \ \ }  \langle F_{T} \rangle_{Z^{\prime}}  =  0.134 ^{+0.012}_{-0.000}\;, 
\end{eqnarray*}
where it can be seen that the value in the $Z^{\prime}$ is very close to the SM result. However, at the low-recoil 
\begin{eqnarray*}
	\langle F_{T} \rangle_{\text{SM}}  =  0.287 ^{+0.008}_{-0.010}\;, \text{ \ \ \ \ \ \ \ \ \ \ \ \ \ \ }  \langle F_{T} \rangle_{Z^{\prime}}  =  0.410 ^{+0.005}_{-0.007}\;,
\end{eqnarray*}
the results of $Z^{\prime}$ model significantly differ from that of the SM. Hence, the measurement of fraction of transverse polarization at low-recoil region will help us to see the possible contribution of neutral $Z^{\prime}$ boson in these $b-$ baryon decays. 

It is well known that in case of the $\Lambda_{b} \to \Lambda J/\psi$ the different asymmetries have been experimentally measured. Motivated by this fact, let's explore the asymmetries corresponding to the hadronic angle $\theta_{\Lambda}$ and $\theta_{l}$ one by one. The asymmetry arising due to the angle $\theta_{\Lambda}$ is defined as $\alpha_{\theta_{\Lambda}}$ and its explicit expression is given in Eq. (\ref{asymmetry1}) and the corresponding numerical values at low- and large-recoil bins are tabulated in Table \ref{low-high-bin-values}. In the large-recoil bin $s \in [1\;, 6]$ GeV$^2$ the value reads as
\begin{eqnarray*}
		\langle \alpha_{\theta_{\Lambda}} \rangle_{\text{SM}}  =  -0.984 ^{+0.007}_{-0.001}\;,  \text{ \ \ \ \ \ \ \ \ \ \ \ \ \ \ }  \langle \alpha_{\theta_{\Lambda}} \rangle_{Z^{\prime}}  =  -0.390 ^{+0.027}_{-0.006}\;.
\end{eqnarray*}
Similarly, in low recoil bin $s \in [15\;,20]$ GeV$^2$, our calculated results for this observable are
\begin{eqnarray*}
	\langle \alpha_{\theta_{\Lambda}} \rangle_{\text{SM}}  =  -0.851 ^{+0.010}_{-0.007}\;, \text{ \ \ \ \ \ \ \ \ \ \ \ \ \ \ }  \langle \alpha_{\theta_{\Lambda}} \rangle_{Z^{\prime}}  =  -0.427 ^{+0.001}_{-0.001}\;.
\end{eqnarray*}
Here we can see that $\alpha_{\theta_{\Lambda}}$ differs in $Z^{\prime}$ model from the SM results significantly in both low- and large-recoil bins. 

Likewise, the asymmetry $\alpha^{\prime}_{\theta_{\ell}}$ that correspond to angle $\theta_{\ell}$ given in Eq. (\ref{asymmetry2}) depends on the angular coefficient $K_{1c}$, therefore, its behavior is similar to $A^{\ell}_{FB}$. Results in the large-recoil bin $s \in [1,6]$ GeV$^2$ for the SM and $Z^{\prime}$ model are
\begin{eqnarray*}
	\langle \alpha^{\prime}_{\theta_{\ell}} \rangle_{\text{SM}}  =  0.047 ^{+0.039}_{-0.016}\;, \text{ \ \ \ \ \ \ \ \ \ \ \ \ \ \ }  \langle \alpha^{\prime}_{\theta_{\ell}} \rangle_{Z^{\prime}}  =  0.027 ^{+0.001}_{-0.002}\;,
\end{eqnarray*}
where the value of $\alpha^{\prime}_{\theta_{\ell}}$ in the SM is $1.7$ times to that of the $Z^{\prime}$ model value. Similarly in low-recoil bin $(s\in [15, 20]$ GeV$^2)$ the results are
\begin{eqnarray*}
	\langle \alpha^{\prime}_{\theta_{\ell}} \rangle_{\text{SM}}  =  -0.280 ^{+0.012}_{-0.010}\;, \text{ \ \ \ \ \ \ \ \ \ \ \ \ \ \ }  \langle \alpha^{\prime}_{\theta_{\ell}} \rangle_{Z^{\prime}}  =  -0.225 ^{+0.006}_{-0.004}\;.
\end{eqnarray*}
It can be noticed that the results in low-recoil bin are an order of magnitude large than the corresponding values in the large-recoil bin both in the SM and in $Z^{\prime}$ model. These values are quite large to be measured at the LCHb experiment to test the predictions of the SM.

We now discuss $\alpha_{\theta_{\ell}}$, which depends upon the angular coefficients $K_{1ss}$ and $K_{1cc}$. This is not significantly affected from the couplings of $Z^{\prime}$ model 
 and hence show little deviations from the SM especially in the large-recoil region. In this region the numerical values are
\begin{eqnarray*}
	\langle \alpha_{\theta_{\ell}} \rangle_{\text{SM}}  =  -0.854 ^{+0.024}_{-0.002}\;, \text{ \ \ \ \ \ \ \ \ \ \ \ \ \ \ }  \langle \alpha_{\theta_{\ell}} \rangle_{Z^{\prime}}  =  -0.857 ^{+0.014}_{-0.001},
\end{eqnarray*}
where it is clear that the values in both the models are almost the same. Similarly in low-recoil region, the results in the SM and $Z^{\prime}$ model are
\begin{eqnarray*}
	\langle \alpha_{\theta_{\ell}} \rangle_{\text{SM}}  =  -0.665 ^{+0.010}_{-0.014}\;, \text{ \ \ \ \ \ \ \ \ \ \ \ \ \ \ }  \langle \alpha_{\theta_{\ell}} \rangle_{Z^{\prime}}  =  -0.485 ^{+0.008}_{-0.011}.
\end{eqnarray*}
In comparison to the low $s$ region, here the values of  $\alpha^{\prime}_{\theta_{\ell}}$ in the SM and $Z^{\prime}$ model differs significantly. Therefore, to establish the possible new physics arising in the $Z^{\prime}$ model, the analysis of $\alpha_{\theta_{\ell}}$ in the high $s$ region will serve as a useful probe. 

Having a look at $\alpha_{\phi}$ discloses that it depends upon $K_{4s}$. At very low $s$, $C_{7}$ term dominates in the SM which results in negative $K_{4s}$ but for $s>2\;$ GeV$^{2}$, the Wilson Coefficient $C_9$ term dominates which give positive results. But for $Z^{\prime}$ model $C^{Z^{\prime}}_{9}$ get affected much more than $C^{Z^{\prime}}_{7}$ for the entire range of $s$ and hence $\alpha_{\phi}$ is expected to be changed significantly with $s$ in the $Z^{\prime}$ model from the corresponding SM result. The values of $\alpha_{\phi}$ in the bin $s\in [1,6]$ GeV$^{2}$ for the SM and $Z^{\prime}$ model are
\begin{eqnarray*}
	\langle \alpha_{\phi} \rangle_{\text{SM}}  =  0.040 ^{+0.070}_{-0.016} \text{ \ \ \ \ \ \ \ \ \ \ \ \ \ \ }  \langle \alpha_{\phi} \rangle_{Z^{\prime}}  =  0.130 ^{+0.015}_{-0.060}
\end{eqnarray*}
and for low recoil region $s\in[15, 20]$ GeV$^{2}$, the values of observable are
\begin{eqnarray*}
	\langle \alpha_{\phi} \rangle_{\text{SM}}  =  0.047 ^{+0.003}_{-0.004}\;, \text{ \ \ \ \ \ \ \ \ \ \ \ \ \ \ }  \langle \alpha_{\phi} \rangle_{Z^{\prime}}  =  -0.448 ^{+0.004}_{-0.006}\;.
\end{eqnarray*}
Hence, it can be revealed that in the SM the value of $\alpha_{\phi}$ is almost the same in low- and large-recoil bins, which is not the case for the  $Z^{\prime}$ model where a large deviation is observed in both the bins. Also in both these bins the results of $Z^{\prime}$ model are quite large compared to the SM results and the experimental observation of $\alpha_{\phi}$ will act as a useful observable.

The longitudinal (transverse) asymmetry parameter $\alpha_{L}$ $(\alpha_{U})$ is ratio of the helicity combinations $K_{2ss}$ $(K_{2cc})$ to  $K_{1ss}$ as depicted in Eq. (\ref{Long-Trans-asymmetry}). Their value in the large-recoil region is
\begin{eqnarray*}
	\langle \alpha_{L} (\alpha_{U}) \rangle_{\text{SM}}  =-0.989 ^{+0.006}_{-0.000}     (-0.916^{+0.010}_{-0.004}) \text{ \ \ \ \ \ \ \ \ \ \ \ \ \ \ }  \langle \alpha_{L} (\alpha_{U}) \rangle_{Z^{\prime}}  =-0.386^{+0.016}_{-0.003}  (-0.445 ^{+0.168}_{-0.040})
\end{eqnarray*}
where we can see that in this bin the values of both the longitudinal and the transverse asymmetry parameters in $Z^{\prime}$ model differ significantly from their respective values in the SM. This is due to the fact that the contribution of extra neutral boson $Z^{\prime}$ affects the value of    
$K_{1ss}$ lesser than the $K_{2ss}(K_{2cc})$.
Now in the low-recoil region
\begin{eqnarray*}
	\langle \alpha_{L} (\alpha_{U}) \rangle_{\text{SM}}  =-0.852 ^{+0.011}_{-0.008}  (-0.844 ^{+0.003}_{-0.002}) \text{ \ \ \ \ \ \ \ \ \ \ \ \ \ \ }  \langle \alpha_{L} (\alpha_{U}) \rangle_{Z^{\prime}}  =-0.458 ^{+0.001}_{-0.001}  (-0.307 ^{+0.002}_{-0.002}).
\end{eqnarray*}
It can be deduced that the value of $\alpha_{L}(\alpha_{U})$ in the $Z^{\prime}$ is half to that of the SM value in this bin. With the current luminosity of the LHCb experiment, the value of these observables is in the measurable range.  Hence the experimental observation of these observables will give us a chance to test the predictions of the SM and a possibility to explore the imprints of $Z^{\prime}$ boson in $\Lambda_{b}\to \Lambda \mu^{+}\mu^{-}$ decays.

It is a well established fact that certain asymmetries, such as $\mathcal{P}^{(\prime)}_{5}$ , that correspond to different foldings in $B \to K^* \mu^{+} \mu^{-}$ have shown significant deviations from the SM predictions. This make them a fertile hunting ground to dig for the various beyond SM scenarios that give possible explanation and $Z^{\prime}$ is one of them \cite{AR-IA}. Motivated by this fact, we have calculated such foldings in the decay under consideration and their expressions in terms of the helicity combinations are given in Eq. (\ref{various-pasym}). Among them the first one is the  $\mathcal{P}_1$ which behaves very similar as $F_T$. The average values of $ \mathcal{P}_1$ at large-recoil in the SM and $Z^{\prime}$ model are
\begin{eqnarray*}
	\langle  \mathcal{P}_1 \rangle_{\text{SM}}  =0.796 ^{+0.002}_{-0.031}\;,      \text{ \ \ \ \ \ \ \ \ \ \ \ \ \ \ }  \langle \mathcal{P}_1 \rangle_{Z^{\prime}}  =0.799^{+0.001}_{-0.018}
\end{eqnarray*}
and at low-recoil, the values are turn out to be
\begin{eqnarray*}
	\langle \mathcal{P}_1 \rangle_{\text{SM}}  =0.569 ^{+0.017}_{-0.009}\;,   \text{ \ \ \ \ \ \ \ \ \ \ \ \ \ \ }  \langle \mathcal{P}_1 \rangle_{Z^{\prime}}  =0.386 ^{+0.010}_{-0.008}\;.
\end{eqnarray*}
Here, we can see that in the large-recoil region, the values in the SM and $Z^{\prime}$ model are very close which is not the situation in the low-recoil region where the value of the SM is $1.5$ times to that of the $Z^{\prime}$ model.

 $ \mathcal{P}_2$ is the ratio of linear combination of $K_{2ss}$ and $K_{2cc}$ to total decay rate. In most of the bins the SM results are more than twice of $Z^{\prime}$ model values and this can be seen in the results at large-recoil, which are
\begin{eqnarray*}
	\langle  \mathcal{P}_2 \rangle_{\text{SM}}  =0.512 ^{+0.001}_{-0.022}\;,      \text{ \ \ \ \ \ \ \ \ \ \ \ \ \ \ }  \langle \mathcal{P}_2 \rangle_{Z^{\prime}}  =0.193^{+0.001}_{-0.002}\;.
\end{eqnarray*}
Similarly the situation persists at the low-recoil
\begin{eqnarray*}
	\langle \mathcal{P}_2 \rangle_{\text{SM}}  =0.316 ^{+0.003}_{-0.002}\;,   \text{ \ \ \ \ \ \ \ \ \ \ \ \ \ \ }  \langle \mathcal{P}_2 \rangle_{Z^{\prime}}  =0.153 ^{+0.004}_{-0.003}\;.
\end{eqnarray*} 

The behavior of $ \mathcal{P}_3$ is similar to $A^{\Lambda}_{FB}$. The average values of $ \mathcal{P}_3$ at large-recoil are
\begin{eqnarray*}
	\langle  \mathcal{P}_3 \rangle_{\text{SM}}  =-0.030 ^{+0.009}_{-0.015}\;,      \text{ \ \ \ \ \ \ \ \ \ \ \ \ \ \ }  \langle \mathcal{P}_3 \rangle_{Z^{\prime}}  =-0.025^{+0.004}_{-0.009}\;,
\end{eqnarray*}
whereas, the results at low-recoil become
\begin{eqnarray*}
	\langle \mathcal{P}_3 \rangle_{\text{SM}}  =0.184 ^{+0.004}_{-0.007}   \text{ \ \ \ \ \ \ \ \ \ \ \ \ \ \ }  \langle \mathcal{P}_3 \rangle_{Z^{\prime}}  =0.232 ^{+0.003}_{-0.004}
\end{eqnarray*}
It can be observed that just like $\mathcal{P}_1$ the asymmetry defined by $\mathcal{P}_3$, the average values in the SM and $Z^{\prime}$ model are comparable at large-recoil but differ significantly at low-recoil. We have observed that with $3$fb$^{-1}$ of data, the LHCb collaboration has measured the $A_{FB}^{h}$ which is of the same order as that of $\mathcal{P}_3$. Therefore, it is expected that in future $\mathcal{P}_3$ will be measured.

Average values of $ \mathcal{P}_5$ at large recoil read as
\begin{eqnarray*}
	\langle  \mathcal{P}_5 \rangle_{\text{SM}}  =0.030 ^{+0.048}_{-0.013}\;,      \text{ \ \ \ \ \ \ \ \ \ \ \ \ \ \ }  \langle \mathcal{P}_5 \rangle_{Z^{\prime}}  =0.034^{+0.013}_{-0.004}\;,
\end{eqnarray*}
and results at low-recoil are
\begin{eqnarray*}
	\langle \mathcal{P}_5 \rangle_{\text{SM}}  =0.163 ^{+0.001}_{-0.000}   \text{ \ \ \ \ \ \ \ \ \ \ \ \ \ \ }  \langle \mathcal{P}_5 \rangle_{Z^{\prime}}  =0.091 ^{+0.002}_{-0.001}
\end{eqnarray*}
The case is also similar to $\mathcal{P}_1$ and $\mathcal{P}_3$ as values in both models are very close at large-recoil and deviations started to appear at low-recoil region of $s$.

Now we come to $\mathcal{P}_6$ which depends on the angular coefficient $K_{4s}$ and hence it behaves as $\alpha_{\phi}$. Values of the observable in the SM and $Z^{\prime}$ at large-recoil become
\begin{eqnarray*}
	\langle  \mathcal{P}_6 \rangle_{\text{SM}}  =0.056 ^{+0.097}_{-0.023}\;,      \text{ \ \ \ \ \ \ \ \ \ \ \ \ \ \ }  \langle \mathcal{P}_6 \rangle_{Z^{\prime}}  =0.180^{+0.021}_{-0.083}\;,
\end{eqnarray*}
and results at low-recoil are
\begin{eqnarray*}
	\langle \mathcal{P}_6 \rangle_{\text{SM}}  =0.066 ^{+0.002}_{-0.007}\;,   \text{ \ \ \ \ \ \ \ \ \ \ \ \ \ \ }  \langle \mathcal{P}_6 \rangle_{Z^{\prime}}  =0.621 ^{+0.005}_{-0.008}\;.
\end{eqnarray*}
From above results, it can be easily extracted that the value of  $\mathcal{P}_6$ in $Z^{\prime}$ model differs significantly from the SM results both at large- and low-recoil which is also the case for $\alpha_{\phi}$. Especially, in the low-recoil region, the value of an asymmetry is an order of magnitude larger from that in the large-recoil bin and it is in the experimentally measurable range with the current luminosity of the LHCb experiment.

The next observable to be discussed is $\mathcal{P}_8$ which mainly depends on the angular coefficient $K_{1c}$ and therefore its behavior is exactly similar as $A^{l}_{FB}$. Its results in large-recoil bin are
\begin{eqnarray*}
	\langle  \mathcal{P}_8 \rangle_{\text{SM}}  =0.088 ^{+0.070}_{-0.032}\;,      \text{ \ \ \ \ \ \ \ \ \ \ \ \ \ \ }  \langle \mathcal{P}_8 \rangle_{Z^{\prime}}  =0.051^{+0.003}_{-0.003}\;,
\end{eqnarray*}
and at low-recoil
\begin{eqnarray*}
	\langle \mathcal{P}_8 \rangle_{\text{SM}}  =-0.480 ^{+0.020}_{-0.012}\;,   \text{ \ \ \ \ \ \ \ \ \ \ \ \ \ \ }  \langle \mathcal{P}_8 \rangle_{Z^{\prime}}  =-0.359 ^{+0.007}_{-0.006}
\end{eqnarray*}
We can see that there is an order of magnitude difference between the results at the large- and low-recoil region. Therefore, the number of events required to see the deviations in the low-recoil region are much smaller compared to dig out the results of this asymmetry in large-recoil region.

The last observable in this list is $\mathcal{P}_9$ which depends upon the angular coefficient $K_{2cc}$. Its values at large-recoil are
\begin{eqnarray*}
	\langle  \mathcal{P}_9 \rangle_{\text{SM}}  =-0.160 ^{+0.001}_{-0.023}\;,      \text{ \ \ \ \ \ \ \ \ \ \ \ \ \ \ }  \langle \mathcal{P}_9 \rangle_{Z^{\prime}}  =-0.076^{+0.024}_{-0.007}\;,
\end{eqnarray*}
and at low-recoil, the results become
\begin{eqnarray*}
	\langle \mathcal{P}_9 \rangle_{\text{SM}}  =-0.308 ^{+0.013}_{-0.008}\;,   \text{ \ \ \ \ \ \ \ \ \ \ \ \ \ \ }  \langle \mathcal{P}_9 \rangle_{Z^{\prime}}  =-0.161 ^{+0.003}_{-0.004}\;.
\end{eqnarray*}
We can see that the value of the SM is almost twice to that of the $Z^{\prime}$ model in both regions.

\begin{figure}
\begin{center}
\begin{tabular}{ll}
	\\ \\	\ \ \ \ \ \ \
	\includegraphics[height=6cm,width=8cm]{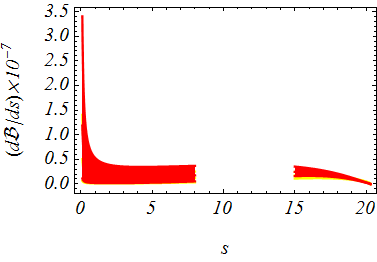} & \ \ \ \includegraphics[height=6.2cm,width=8.5cm]{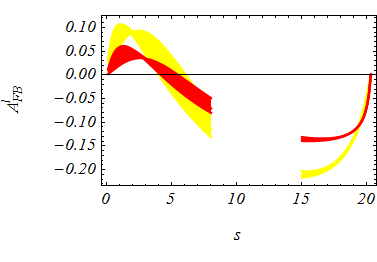} \\ 
	\includegraphics[height=6.5cm,width=8.7cm]{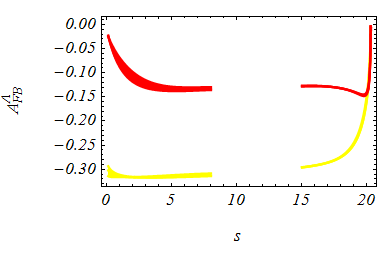} & \ \ \ \includegraphics[height=6.5cm,width=8.5cm]{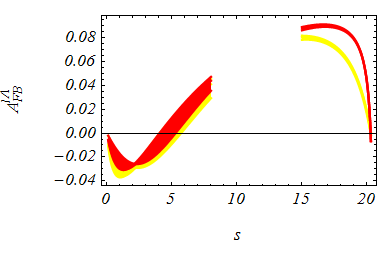}\\
	\ \ \ \ \ \ \ \ \
	\includegraphics[height=6cm,width=7.7cm]{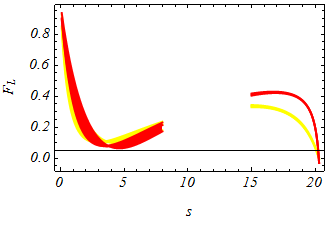} & 
	\ \ \ \ \ \ \ \ \ \ \ \
    \includegraphics[height=6cm,width=7.6cm]{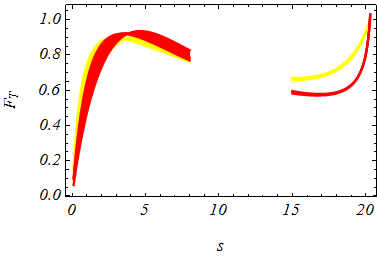}
\end{tabular}
\end{center}
\caption{Branching ratio and various forward-backward asymmetries are plotted as a function of $s$. The yellow curve correspond the SM results and red to the  $Z^{\prime}$ model. In both cases, the bands corresponds to the uncertainties in the form factors and other input parameters. }\label{Fig2}
\end{figure}

\begin{figure}
	\begin{center}
		\begin{tabular}{ll}
			\\ \\
			 \includegraphics[height=6.5cm,width=8.5cm]{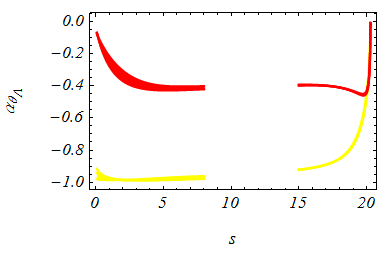} & \ \ \
			\includegraphics[height=6.5cm,width=8.5cm]{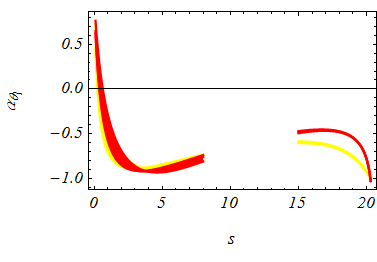} \\  \includegraphics[height=6.5cm,width=8.5cm]{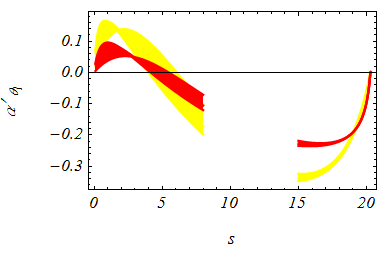}& \ \ \
			\includegraphics[height=6.5cm,width=8.5cm]{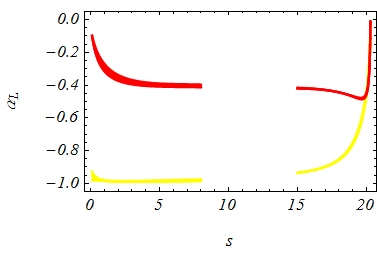}\\ \  \includegraphics[height=6.7cm,width=8.5cm]{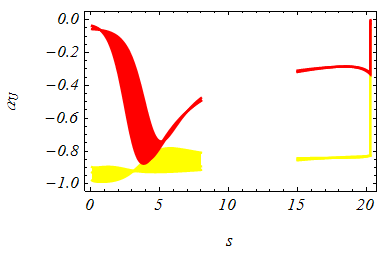} & \ \ \ \ \ \ \ \ \
			 \includegraphics[height=6.4cm,width=7.8cm]{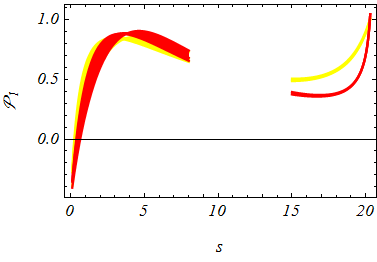}
		\end{tabular}
	\end{center}
	\caption{Different asymmetry parameters denoted by $\alpha$ and $\mathcal{P}_1$ are plotted as function of $s$. The color coding is same as in Fig. \ref{Fig2}.}\label{Fig1}
\end{figure}

\begin{figure}
	\begin{center}
		\begin{tabular}{ll}
			\\ \\ 	\ \ \ \includegraphics[height=6.3cm,width=8.5cm]{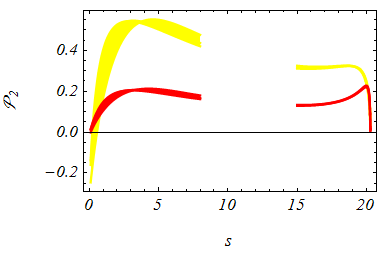} & \ \ \ \includegraphics[height=6.5cm,width=8.5cm]{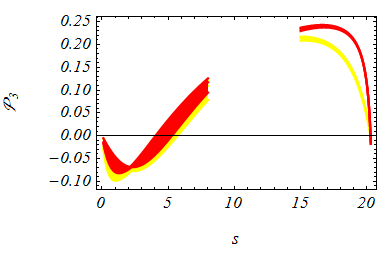}\\
			\ \ \ \ \	\includegraphics[height=6.3cm,width=8.3cm]{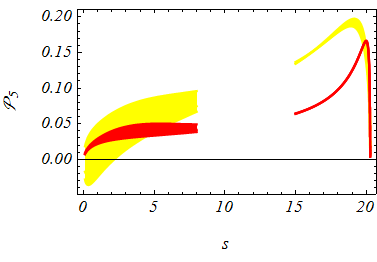} & \ \ \ \ \ \  \includegraphics[height=6.3cm,width=8.1cm]{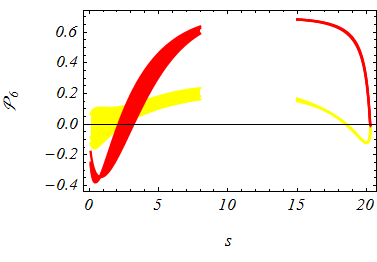}\\
			\ \ \ \  \includegraphics[height=6.5cm,width=8.5cm]{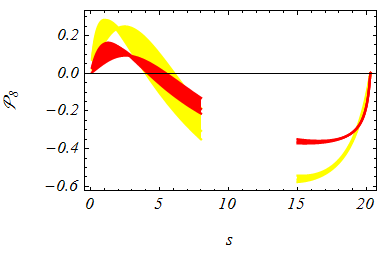} & \ \ \ \ \ \ \includegraphics[height=6.5cm,width=8.2cm]{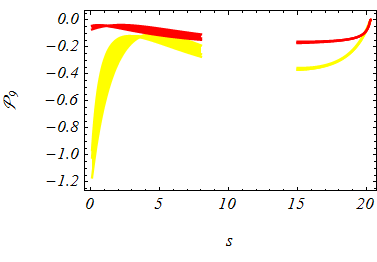}
		\end{tabular}
	\end{center}
	\caption{The folded distributions $\mathcal{P}_{2, \cdots, 9}$, except $\mathcal{P}_4$ are plotted as function of $s$. The color coding is same as in Fig. \ref{Fig2}.}\label{Fig3}
\end{figure}

In case of $\Lambda_b \to \Lambda (\to p\pi)\mu^{+}\mu^{-}$ decay, the LHCb experiment has measured the value of branching ratio, forward-backward asymmetries and longitudinal dimuon helicity fraction in small bins of $s$. Therefore, we have tabulated the values of above mentioned observables in large- and low-recoil region in Table \ref{low-high-bin-values} and various small bins in Tables \ref{Bin-analysis} and \ref{pbin-values}. In addition, to see the profile of these asymmetries we have plotted them graphically in Figs. \ref{Fig2}, \ref{Fig1} and Fig. \ref{Fig3} with the square of dimuon momentum $s$. We hope that in future when more precise results of various asymmetries will come from the LHCb, it will give us a chance to compare the profile of various asymmetries calculated here with the experiments both for the SM and $Z^{\prime}$ model.

%
\begin{table}
\caption{Average values of different observables for $\Lambda_b \to \Lambda (\to p\pi)\mu^{+}\mu^{-}$ in low and large recoil regions.}\label{low-high-bin-values}
\begin{tabular}{l||llllll}
\hline\hline
& $\ \ \ \ \ \ \left\langle \alpha _{\theta _{\Lambda }}\right\rangle $ & $\
\ \ \ \ \ \left\langle \alpha _{\theta _{l}}^{\prime }\right\rangle $ & $\ \
\ \ \ \ \left\langle \alpha _{\theta _{l}}\right\rangle $ & $\ \ \ \ \ \
\left\langle \alpha _{\phi }\right\rangle $ & $\ \ \ \ \ \ \left\langle
\alpha _{\phi }^{\prime }\right\rangle $ & $\ \ \ \ \ \ \left\langle \alpha
_{U}\right\rangle $ \\ \hline
$\lbrack 1,6]%
\begin{tabular}{l}
$SM$ \\ 
$Z^\prime $%
\end{tabular}%
$ & $%
\begin{tabular}{l}
$-0.984_{-0.001}^{+0.007}$ \\ 
$-0.390_{-0.006}^{+0.027}$%
\end{tabular}%
$ & $%
\begin{tabular}{l}
$ \ \ 0.047_{-0.016}^{+0.039}$ \\ 
$ \ \ 0.027_{-0.002}^{+0.001}$%
\end{tabular}%
$ & $%
\begin{tabular}{l}
$-0.854_{-0.002}^{+0.024}$ \\ 
$-0.857_{-0.001}^{+0.014}$%
\end{tabular}%
$ & $%
\begin{tabular}{l}
$ \ \ 0.040_{-0.016}^{+0.070}$ \\ 
$ \ \ 0.130_{-0.060}^{+0.015}$%
\end{tabular}%
$ & $%
\begin{tabular}{l}
$ \ \ 0.000_{-0.001}^{+0.000}$ \\ 
$-0.002_{-0.000}^{+0.001}$%
\end{tabular}%
$ & $%
\begin{tabular}{l}
$-0.916_{-0.004}^{+0.010}$ \\ 
$-0.445_{-0.040}^{+0.168}$%
\end{tabular}%
$ \\ \hline
$\lbrack 15,20.25]%
\begin{tabular}{l}
$SM$ \\ 
$Z^\prime $%
\end{tabular}%
$ & $%
\begin{tabular}{l}
$-0.851_{-0.007}^{+0.010}$ \\ 
$-0.427_{-0.001}^{+0.001}$%
\end{tabular}%
$ & $%
\begin{tabular}{l}
$-0.280_{-0.010}^{+0.012}$ \\ 
$-0.225_{-0.004}^{+0.006}$%
\end{tabular}%
$ & $%
\begin{tabular}{l}
$-0.665_{-0.014}^{+0.010}$ \\ 
$-0.485_{-0.011}^{+0.008}$%
\end{tabular}%
$ & $%
\begin{tabular}{l}
$ \ \ 0.047_{-0.004}^{+0.003}$ \\ 
$ \ \ 0.448_{-0.006}^{+0.004}$%
\end{tabular}%
$ & $%
\begin{tabular}{l}
$-0.056_{-0.002}^{+0.001}$ \\ 
$-0.049_{-0.002}^{+0.002}$%
\end{tabular}%
$ & $%
\begin{tabular}{l}
$-0.844_{-0.002}^{+0.003}$ \\ 
$-0.307_{-0.002}^{+0.002}$%
\end{tabular}%
$ \\ \hline\hline
& $\ \ \ \ \ \ \left\langle \alpha _{L}\right\rangle $ & $\left\langle
d\mathcal{B}/ds\right\rangle \times 10^{-7}$ & $\ \ \ \left\langle
F_{T}\right\rangle $ & $\ \ \ \left\langle F_{L}\right\rangle $ & $\ \ \ \ \
\ \left\langle A_{FB}^{\ell\Lambda }\right\rangle $ & $\ \ \ \ \ \ \left\langle
\mathcal{P}_{1}\right\rangle $ \\ \hline
$\lbrack 1,6]\ 
\begin{tabular}{l}
$SM$ \\ 
$Z^\prime $%
\end{tabular}%
$ & $%
\begin{tabular}{l}
$-0.989_{-0.000}^{+0.006}$ \\ 
$-0.386_{-0.003}^{+0.016}$%
\end{tabular}%
$ & $%
\begin{tabular}{l}
$ \ \ 0.466_{-0.394}^{+0.760}$ \\ 
$ \ \ 0.709_{-0.601}^{+0.115}$%
\end{tabular}%
$ & $%
\begin{tabular}{l}
$ \ \ 0.136_{-0.002}^{+0.021}$ \\ 
$ \ \ 0.134_{-0.000}^{+0.012}$%
\end{tabular}%
$ & $%
\begin{tabular}{l}
$ \ \ 0.864_{-0.021}^{+0.002}$ \\ 
$ \ \ 0.866_{-0.012}^{+0.000}$%
\end{tabular}%
$ & $%
\begin{tabular}{l}
$-0.011_{-0.006}^{+0.003}$ \\ 
$-0.009_{-0.003}^{+0.002}$%
\end{tabular}%
$ & $%
\begin{tabular}{l}
$ \ \ 0.796_{-0.031}^{+0.002}$ \\ 
$ \ \ 0.799_{-0.018}^{+0.001}$%
\end{tabular}%
$ \\ \hline
$\lbrack 15,20.25]%
\begin{tabular}{l}
$SM$ \\ 
$Z^\prime $%
\end{tabular}%
$ & $%
\begin{tabular}{l}
$-0.852_{-0.008}^{+0.011}$ \\ 
$-0.458_{-0.001}^{+0.001}$%
\end{tabular}%
$ & $%
\begin{tabular}{l}
$ \ \ 0.731_{0.187}^{0.198}$ \\ 
$ \ \ 1.179_{0.233}^{0.271}$%
\end{tabular}%
$ & $%
\begin{tabular}{l}
$ \ \ 0.287_{-0.010}^{+0.008}$ \\ 
$ \ \ 0.410_{-0.007}^{+0.005}$%
\end{tabular}%
$ & $%
\begin{tabular}{l}
$ \ \ 0.713_{-0.008}^{+0.010}$ \\ 
$ \ \ 0.590_{-0.005}^{+0.007}$%
\end{tabular}%
$ & $%
\begin{tabular}{l}
$ \ \ 0.069_{-0.002}^{+0.002}$ \\ 
$ \ \ 0.087_{-0.002}^{+0.001}$%
\end{tabular}%
$ & $%
\begin{tabular}{l}
$ \ \ 0.569_{-0.009}^{+0.017}$ \\ 
$ \ \ 0.386_{-0.008}^{+0.010}$%
\end{tabular}%
$%
\end{tabular}
\begin{tabular}{l||llllll}
\hline\hline
& $\ \ \ \ \ \ \ \left\langle \mathcal{P}_{2}\right\rangle $ & $\ \ \ \ \ \ \left\langle
\mathcal{P}_{3}\right\rangle $ & $\ \ \ \ \ \ \left\langle \mathcal{P}_{5}\right\rangle $ & $\ \
\ \ \ \ \left\langle \mathcal{P}_{6}\right\rangle $ & $\ \ \ \ \ \ \left\langle
\mathcal{P}_{8}\right\rangle $ & $\ \ \ \ \ \ \left\langle \mathcal{P}_{9}\right\rangle $ \\ 
\hline
$\lbrack 1,6]%
\begin{tabular}{l}
$SM$ \\ 
$Z^\prime $%
\end{tabular}%
$ & $%
\begin{tabular}{l}
$ \ \ 0.512_{-0.022}^{+0.001}$ \\ 
$ \ \ 0.193_{-0.002}^{+0.001}$%
\end{tabular}%
$ & $%
\begin{tabular}{l}
$-0.030_{-0.015}^{+0.009}$ \\ 
$-0.025_{-0.009}^{+0.004}$%
\end{tabular}%
$ & $%
\begin{tabular}{l}
$ \ \ 0.030_{-0.013}^{+0.048}$ \\ 
$ \ \ 0.034_{-0.004}^{+0.013}$%
\end{tabular}%
$ & $%
\begin{tabular}{l}
$ \ \ 0.056_{-0.023}^{+0.097}$ \\ 
$ \ \ 0.180_{-0.083}^{+0.021}$%
\end{tabular}%
$ & $%
\begin{tabular}{l}
$ \ \ 0.088_{-0.032}^{+0.070}$ \\ 
$ \ \ 0.051_{-0.003}^{+0.003}$%
\end{tabular}%
$ & $%
\begin{tabular}{l}
$-0.160_{-0.023}^{+0.001}$ \\ 
$-0.076_{-0.007}^{+0.024}$%
\end{tabular}%
$ \\ \hline
$\lbrack 15,20.25]%
\begin{tabular}{l}
$SM$ \\ 
$Z^\prime $%
\end{tabular}%
$ & $%
\begin{tabular}{l}
$ \ \ 0.316_{-0.002}^{+0.003}$ \\ 
$ \ \ 0.153_{-0.003}^{+0.004}$%
\end{tabular}%
$ & $%
\begin{tabular}{l}
$ \ \ 0.184_{-0.007}^{+0.004}$ \\ 
$ \ \ 0.232_{-0.004}^{+0.003}$%
\end{tabular}%
$ & $%
\begin{tabular}{l}
$ \ \ 0.163_{-0.000}^{+0.001}$ \\ 
$ \ \ 0.091_{-0.001}^{+0.002}$%
\end{tabular}%
$ & $%
\begin{tabular}{l}
$ \ \ 0.066_{-0.007}^{+0.002}$ \\ 
$ \ \ 0.621_{-0.008}^{+0.005}$%
\end{tabular}%
$ & $%
\begin{tabular}{l}
$-0.480_{-0.012}^{+0.020}$ \\ 
$-0.359_{-0.006}^{+0.007}$%
\end{tabular}%
$ & $%
\begin{tabular}{l}
$-0.308_{-0.008}^{+0.013}$ \\ 
$-0.161_{-0.003}^{+0.004}$%
\end{tabular}%
$%
\end{tabular}
\end{table}

\begin{table}
		\caption{Numerical results for observables for the decay $\Lambda_b \rightarrow \Lambda (\rightarrow p \pi) \mu^+ \mu^-$ for the SM and $Z^{\prime}$ in appropriate bins.}\label{Bin-analysis}
\begin{tabular}{l||llllll}
\hline\hline
& $\ \ \ \ \ \ \left\langle \alpha _{\theta _{\Lambda }}\right\rangle $ & $\
\ \ \ \ \ \left\langle \alpha _{\theta _{l}}^{\prime }\right\rangle $ & $\ \ \ \ \ \left\langle \alpha _{\theta _{l}}\right\rangle $ & $\ \ \ \ \ \
\left\langle \alpha _{\phi }\right\rangle $ & $\ \ \ \ \ \ \left\langle
\alpha _{\phi }^{\prime }\right\rangle $ & $\ \ \ \ \ \ \left\langle \alpha
_{U}\right\rangle $ \\ \hline\hline
$\lbrack 0.1,2]%
\begin{tabular}{l}
$SM$ \\ 
$Z^\prime $%
\end{tabular}%
$ & $%
\begin{tabular}{l}
$-0.970_{-0.014}^{+0.012}$ \\ 
$-0.209_{-0.005}^{+0.028}$%
\end{tabular}%
$ & $%
\begin{tabular}{l}
$ \ \ 0.140_{-0.049}^{+0.001}$ \\ 
$ \ \ 0.073_{-0.042}^{+0.005}$%
\end{tabular}%
$ & $%
\begin{tabular}{l}
$-0.463_{-0.049}^{+0.315}$ \\ 
$-0.265_{-0.033}^{+0.191}$%
\end{tabular}%
$ & $%
\begin{tabular}{l}
$-0.058_{-0.026}^{+0.129}$ \\ 
$-0.208_{-0.001}^{+0.003}$%
\end{tabular}%
$ & $%
\begin{tabular}{l}
$ \ \ 0.002_{-0.001}^{+0.000}$ \\ 
$ \ \ 0.000_{-0.000}^{+0.000}$%
\end{tabular}%
$ & $%
\begin{tabular}{l}
$-0.933_{-0.051}^{+0.034}$ \\ 
$-0.063_{-0.001}^{+0.001}$%
\end{tabular}%
$ \\ \hline
$\lbrack 1,2]%
\begin{tabular}{l}
$SM$ \\ 
$Z^\prime $%
\end{tabular}%
$ & $%
\begin{tabular}{l}
$-0.983_{-0.002}^{+0.005}$ \\ 
$-0.318_{-0.010}^{+0.054}$%
\end{tabular}%
$ & $%
\begin{tabular}{l}
$ \ \ 0.141_{-0.013}^{+0.008}$ \\ 
$\ \ 0.086_{-0.040}^{+0.003}$%
\end{tabular}%
$ & $%
\begin{tabular}{l}
$-0.788_{-0.028}^{+0.223}$ \\ 
$-0.710_{-0.030}^{+0.193}$%
\end{tabular}%
$ & $%
\begin{tabular}{l}
$-0.032_{-0.020}^{+0.110}$ \\ 
$-0.137_{-0.074}^{+0.020}$%
\end{tabular}%
$ & $%
\begin{tabular}{l}
$ \ \ 0.001_{-0.001}^{+0.000}$ \\ 
$ \ \ 0.000_{-0.001}^{+0.000}$%
\end{tabular}%
$ & $%
\begin{tabular}{l}
$-0.932_{-0.051}^{+0.036}$ \\ 
$-0.140_{-0.024}^{+0.061}$%
\end{tabular}%
$ \\ \hline
$\lbrack 2,4]%
\begin{tabular}{l}
$SM$ \\ 
$Z^\prime $%
\end{tabular}%
$ & $%
\begin{tabular}{l}
$-0.987_{-0.001}^{+0.006}$ \\ 
$-0.397_{-0.008}^{+0.040}$%
\end{tabular}%
$ & $%
\begin{tabular}{l}
$ \ \ 0.076_{-0.020}^{+0.048}$ \\ 
$ \ \ 0.046_{-0.004}^{+0.001}$%
\end{tabular}%
$ & $%
\begin{tabular}{l}
$-0.887_{-0.001}^{+0.046}$ \\ 
$-0.899_{-0.003}^{+0.048}$%
\end{tabular}%
$ & $%
\begin{tabular}{l}
$ \ \ 0.030_{-0.013}^{+0.064}$ \\ 
$ \ \ 0.100_{-0.120}^{+0.029}$%
\end{tabular}%
$ & $%
\begin{tabular}{l}
$ \ \ 0.000_{-0.001}^{+0.000}$ \\ 
$-0.001_{-0.001}^{+0.000}$%
\end{tabular}%
$ & $%
\begin{tabular}{l}
$-0.923_{-0.005}^{+0.007}$ \\ 
$-0.561_{-0.088}^{+0.333}$%
\end{tabular}%
$ \\ \hline
$\lbrack 4,6]%
\begin{tabular}{l}
$SM$ \\ 
$Z^\prime $%
\end{tabular}%
$ & $%
\begin{tabular}{l}
$-0.984_{-0.002}^{+0.012}$ \\ 
$-0.423_{-0.006}^{+0.023}$%
\end{tabular}%
$ & $%
\begin{tabular}{l}
$-0.030_{-0.028}^{+0.079}$ \\ 
$-0.018_{-0.011}^{+0.028}$%
\end{tabular}%
$ & $%
\begin{tabular}{l}
$-0.858_{-0.035}^{+0.016}$ \\ 
$-0.892_{-0.034}^{+0.014}$%
\end{tabular}%
$ & $%
\begin{tabular}{l}
$ \ \ 0.086_{-0.013}^{+0.046}$ \\ 
$ \ \ 0.308_{-0.083}^{+0.022}$%
\end{tabular}%
$ & $%
\begin{tabular}{l}
$-0.001_{-0.001}^{+0.000}$ \\ 
$-0.002_{-0.001}^{+0.000}$%
\end{tabular}%
$ & $%
\begin{tabular}{l}
$-0.908_{-0.020}^{+0.110}$ \\ 
$-0.754_{-0.008}^{+0.060}$%
\end{tabular}%
$ \\ \hline
$\lbrack 6,8]%
\begin{tabular}{l}
$SM$ \\ 
$Z^\prime $%
\end{tabular}%
$ & $%
\begin{tabular}{l}
$-0.977_{-0.004}^{+0.014}$ \\ 
$-0.420_{-0.005}^{+0.016}$%
\end{tabular}%
$ & $%
\begin{tabular}{l}
$-0.128_{-0.028}^{+0.075}$ \\ 
$-0.078_{-0.014}^{+0.034}$%
\end{tabular}%
$ & $%
\begin{tabular}{l}
$-0.789_{-0.049}^{+0.020}$ \\ 
$-0.809_{-0.050}^{+0.020}$%
\end{tabular}%
$ & $%
\begin{tabular}{l}
$ \ \ 0.200_{-0.014}^{+0.041}$ \\ 
$ \ \ 0.418_{-0.038}^{+0.012}$%
\end{tabular}%
$ & $%
\begin{tabular}{l}
$-0.002_{-0.001}^{+0.000}$ \\ 
$-0.002_{-0.000}^{+0.000}$%
\end{tabular}%
$ & $%
\begin{tabular}{l}
$-0.897_{-0.023}^{+0.100}$ \\ 
$-0.556_{-0.000}^{+0.010}$%
\end{tabular}%
$ \\ \hline
$\lbrack 14,16]%
\begin{tabular}{l}
$SM$ \\ 
$Z^\prime $%
\end{tabular}%
$ & $%
\begin{tabular}{l}
$-0.922_{-0.001}^{+0.002}$ \\ 
$-0.409_{-0.003}^{+0.004}$%
\end{tabular}%
$ & $%
\begin{tabular}{l}
$-0.338_{-0.009}^{+0.014}$ \\ 
$-0.231_{-0.006}^{+0.010}$%
\end{tabular}%
$ & $%
\begin{tabular}{l}
$-0.596_{-0.012}^{+0.008}$ \\ 
$-0.482_{-0.012}^{+0.008}$%
\end{tabular}%
$ & $%
\begin{tabular}{l}
$ \ \ 0.114_{-0.005}^{+0.007}$ \\ 
$ \ \ 0.484_{-0.001}^{+0.000}$%
\end{tabular}%
$ & $%
\begin{tabular}{l}
$-0.038_{-0.000}^{+0.000}$ \\ 
$-0.030_{-0.001}^{+0.000}$%
\end{tabular}%
$ & $%
\begin{tabular}{l}
$-0.851_{-0.006}^{+0.009}$ \\ 
$-0.321_{-0.004}^{+0.006}$%
\end{tabular}%
$ \\ \hline
$\lbrack 16,18]%
\begin{tabular}{l}
$SM$ \\ 
$Z^\prime $%
\end{tabular}%
$ & $%
\begin{tabular}{l}
$-0.889_{-0.001}^{+0.002}$ \\ 
$-0.420_{-0.001}^{+0.001}$%
\end{tabular}%
$ & $%
\begin{tabular}{l}
$-0.312_{-0.005}^{+0.007}$ \\ 
$-0.235_{-0.003}^{+0.004}$%
\end{tabular}%
$ & $%
\begin{tabular}{l}
$-0.627_{-0.008}^{+0.006}$ \\ 
$-0.461_{-0.008}^{+0.006}$%
\end{tabular}%
$ & $%
\begin{tabular}{l}
$ \ \ 0.067_{-0.001}^{+0.001}$ \\ 
$ \ \ 0.462_{-0.002}^{+0.001}$%
\end{tabular}%
$ & $%
\begin{tabular}{l}
$-0.052_{-0.000}^{+0.000}$ \\ 
$-0.044_{-0.000}^{+0.000}$%
\end{tabular}%
$ & $%
\begin{tabular}{l}
$-0.843_{-0.002}^{+0.002}$ \\ 
$-0.304_{-0.002}^{+0.002}$%
\end{tabular}%
$ \\ \hline
$\lbrack 18,20.25]%
\begin{tabular}{l}
$SM$ \\ 
$Z^\prime $%
\end{tabular}%
$ & $%
\begin{tabular}{l}
$-0.747_{-0.008}^{+0.010}$ \\ 
$-0.459_{-0.002}^{+0.002}$%
\end{tabular}%
$ & $%
\begin{tabular}{l}
$-0.195_{-0.004}^{+0.005}$ \\ 
$-0.198_{-0.001}^{+0.002}$%
\end{tabular}%
$ & $%
\begin{tabular}{l}
$-0.767_{-0.007}^{+0.006}$ \\ 
$-0.544_{-0.007}^{+0.006}$%
\end{tabular}%
$ & $%
\begin{tabular}{l}
$-0.023_{-0.003}^{+0.002}$ \\ 
$ \ \ 0.382_{-0.006}^{+0.005}$%
\end{tabular}%
$ & $%
\begin{tabular}{l}
$-0.072_{-0.002}^{+0.002}$ \\ 
$-0.078_{-0.003}^{+0.003}$%
\end{tabular}%
$ & $%
\begin{tabular}{l}
$-0.840_{-0.000}^{+0.000}$ \\ 
$-0.302_{-0.001}^{+0.001}$%
\end{tabular}%
$ \\ \hline\hline
& $\ \ \ \ \ \ \left\langle \alpha _{L}\right\rangle $ & 
\begin{tabular}{l}
\ \	$\left\langle d\mathcal{B}/ds\right\rangle $ \\ 
\ \ \ 	$\times 10^{-7}$%
\end{tabular}
& $\ \ \ \left\langle F_{T}\right\rangle $ & $\ \ \ \left\langle
F_{L}\right\rangle $ & $\ \ \ \ \ \ \left\langle A_{FB}^{\ell\Lambda
}\right\rangle $ & $\ \ \ \ \ \ \left\langle \mathcal{P}_{1}\right\rangle $ \\ 
\hline\hline
$\lbrack 0.1,2]%
\begin{tabular}{l}
$SM$ \\ 
$Z^\prime $%
\end{tabular}%
$ & $%
\begin{tabular}{l}
$-0.980_{-0.004}^{+0.008}$ \\ 
$-0.262_{-0.004}^{+0.027}$%
\end{tabular}%
$ & $%
\begin{tabular}{l}
$ \ \ 0.251_{-0.222}^{+0.451}$ \\ 
$ \ \ 0.479_{-0.430}^{+0.880}$%
\end{tabular}%
$ & $%
\begin{tabular}{l}
$ \ \ 0.424_{-0.031}^{+0.174}$ \\ 
$ \ \ 0.537_{-0.018}^{+0.095}$%
\end{tabular}%
$ & $%
\begin{tabular}{l}
$ \ \ 0.576_{-0.174}^{+0.031}$ \\ 
$ \ \ 0.463_{-0.095}^{+0.018}$%
\end{tabular}%
$ & $%
\begin{tabular}{l}
$-0.028_{-0.001}^{+0.013}$ \\ 
$-0.021_{-0.001}^{+0.008}$%
\end{tabular}%
$ & $%
\begin{tabular}{l}
$ \ \ 0.364_{-0.260}^{+0.047}$ \\ 
$ \ \ 0.194_{-0.143}^{+0.027}$%
\end{tabular}%
$ \\ \hline
$\lbrack 1,2]%
\begin{tabular}{l}
$SM$ \\ 
$Z^\prime $%
\end{tabular}%
$ & $%
\begin{tabular}{l}
$-0.986_{-0.003}^{+0.001}$ \\ 
$-0.344_{-0.005}^{+0.036}$%
\end{tabular}%
$ & $%
\begin{tabular}{l}
$ \ \ 0.095_{-0.084}^{+0.172}$ \\ 
$ \ \ 0.158_{-0.140}^{+0.283}$%
\end{tabular}%
$ & $%
\begin{tabular}{l}
$ \ \ 0.192_{-0.023}^{+0.166}$ \\ 
$ \ \ 0.253_{-0.023}^{+0.136}$%
\end{tabular}%
$ & $%
\begin{tabular}{l}
$ \ \ 0.808_{-0.166}^{+0.023}$ \\ 
$ \ \ 0.747_{-0.136}^{+0.023}$%
\end{tabular}%
$ & $%
\begin{tabular}{l}
$-0.034_{-0.001}^{+0.009}$ \\ 
$-0.030_{-0.000}^{+0.007}$%
\end{tabular}%
$ & $%
\begin{tabular}{l}
$ \ \ 0.711_{-0.248}^{+0.035}$ \\ 
$ \ \ 0.620_{-0.203}^{+0.035}$%
\end{tabular}%
$ \\ \hline
$\lbrack 2,4]%
\begin{tabular}{l}
$SM$ \\ 
$Z^\prime $%
\end{tabular}%
$ & $%
\begin{tabular}{l}
$-0.991_{-0.000}^{+0.006}$ \\ 
$-0.389_{-0.004}^{+0.022}$%
\end{tabular}%
$ & $%
\begin{tabular}{l}
$ \ \ 0.178_{-0.153}^{+0.302}$ \\ 
$ \ \ 0.268_{-0.231}^{+4.490}$%
\end{tabular}%
$ & $%
\begin{tabular}{l}
$ \ \ 0.107_{-0.001}^{+0.040}$ \\ 
$ \ \ 0.096_{-0.003}^{+0.042}$%
\end{tabular}%
$ & $%
\begin{tabular}{l}
$ \ \ 0.893_{-0.040}^{+0.001}$ \\ 
$ \ \ 0.904_{-0.042}^{+0.003}$%
\end{tabular}%
$ & $%
\begin{tabular}{l}
$-0.019_{-0.008}^{+0.004}$ \\ 
$-0.017_{-0.006}^{+0.002}$%
\end{tabular}%
$ & $%
\begin{tabular}{l}
$\ \ 0.839_{-0.061}^{+0.001}$ \\ 
$ \ \ 0.856_{-0.064}^{+0.005}$%
\end{tabular}%
$ \\ \hline
$\lbrack 4,6]%
\begin{tabular}{l}
$SM$ \\ 
$Z^\prime $%
\end{tabular}%
$ & $%
\begin{tabular}{l}
$-0.989_{-0.001}^{+0.008}$ \\ 
$-0.404_{-0.004}^{+0.015}$%
\end{tabular}%
$ & $%
\begin{tabular}{l}
$ \ \ 0.193_{-0.157}^{+0.286}$ \\ 
$ \ \ 0.283_{-0.229}^{+0.041}$%
\end{tabular}%
$ & $%
\begin{tabular}{l}
$ \ \ 0.133_{-0.031}^{+0.014}$ \\ 
$ \ \ 0.102_{-0.031}^{+0.013}$%
\end{tabular}%
$ & $%
\begin{tabular}{l}
$ \ \ 0.867_{-0.014}^{+0.031}$ \\ 
$ \ \ 0.897_{-0.013}^{+0.031}$%
\end{tabular}%
$ & $%
\begin{tabular}{l}
$ \ \ 0.007_{-0.015}^{+0.005}$ \\ 
$ \ \ 0.009_{-0.012}^{+0.004}$%
\end{tabular}%
$ & $%
\begin{tabular}{l}
$ \ \ 0.801_{-0.021}^{+0.047}$ \\ 
$ \ \ 0.846_{-0.019}^{+0.046}$%
\end{tabular}%
$ \\ \hline
$\lbrack 6,8]%
\begin{tabular}{l}
$SM$ \\ 
$Z^\prime $%
\end{tabular}%
$ & $%
\begin{tabular}{l}
$-0.986_{-0.002}^{+0.009}$ \\ 
$-0.407_{-0.004}^{+0.013}$%
\end{tabular}%
$ & $%
\begin{tabular}{l}
$ \ \ 0.220_{-0.164}^{+0.275}$ \\ 
$ \ \ 0.325_{-0.240}^{+0.040}$%
\end{tabular}%
$ & $%
\begin{tabular}{l}
$ \ \ 0.191_{-0.041}^{+0.016}$ \\ 
$ \ \ 0.175_{-0.042}^{+0.016}$%
\end{tabular}%
$ & $%
\begin{tabular}{l}
$ \ \ 0.809_{-0.016}^{+0.041}$ \\ 
$ \ \ 0.825_{-0.016}^{+0.042}$%
\end{tabular}%
$ & $%
\begin{tabular}{l}
$ \ \ 0.031_{-0.013}^{+0.005}$ \\ 
$ \ \ 0.034_{-0.010}^{+0.003}$%
\end{tabular}%
$ & $%
\begin{tabular}{l}
$ \ \ 0.714_{-0.025}^{+0.061}$ \\ 
$ \ \ 0.738_{-0.024}^{+0.063}$%
\end{tabular}%
$ \\ \hline
$\lbrack 14,16]%
\begin{tabular}{l}
$SM$ \\ 
$Z^\prime $%
\end{tabular}%
$ & $%
\begin{tabular}{l}
$-0.936_{-0.001}^{+0.001}$ \\ 
$-0.432_{-0.002}^{+0.004}$%
\end{tabular}%
$ & $%
\begin{tabular}{l}
$ \ \ 0.353_{-0.127}^{+0.153}$ \\ 
$ \ \ 0.720_{-0.178}^{+0.217}$%
\end{tabular}%
$ & $%
\begin{tabular}{l}
$ \ \ 0.336_{-0.008}^{+0.005}$ \\ 
$ \ \ 0.411_{-0.008}^{+0.005}$%
\end{tabular}%
$ & $%
\begin{tabular}{l}
$ \ \ 0.664_{-0.005}^{+0.008}$ \\ 
$ \ \ 0.589_{-0.005}^{+0.008}$%
\end{tabular}%
$ & $%
\begin{tabular}{l}
$ \ \ 0.080_{-0.002}^{+0.001}$ \\ 
$ \ \ 0.088_{-0.002}^{+0.001}$%
\end{tabular}%
$ & $%
\begin{tabular}{l}
$ \ \ 0.496_{-0.006}^{+0.013}$ \\ 
$ \ \ 0.383_{-0.008}^{+0.012}$%
\end{tabular}%
$ \\ \hline
$\lbrack 16,18]%
\begin{tabular}{l}
$SM$ \\ 
$Z^\prime $%
\end{tabular}%
$ & $%
\begin{tabular}{l}
$-0.896_{-0.001}^{+0.002}$ \\ 
$-0.451_{-0.001}^{+0.001}$%
\end{tabular}%
$ & $%
\begin{tabular}{l}
$ \ \ 0.328_{-0.086}^{+0.095}$ \\ 
$ \ \ 0.562_{-0.112}^{+0.129}$%
\end{tabular}%
$ & $%
\begin{tabular}{l}
$ \ \ 0.315_{-0.005}^{+0.004}$ \\ 
$ \ \ 0.424_{-0.005}^{+0.004}$%
\end{tabular}%
$ & $%
\begin{tabular}{l}
$ \ \ 0.685_{-0.004}^{+0.005}$ \\ 
$ \ \ 0.576_{-0.004}^{+0.005}$%
\end{tabular}%
$ & $%
\begin{tabular}{l}
$ \ \ 0.076_{-0.001}^{+0.001}$ \\ 
$ \ \ 0.090_{-0.001}^{+0.001}$%
\end{tabular}%
$ & $%
\begin{tabular}{l}
$ \ \ 0.528_{-0.003}^{+0.011}$ \\ 
$ \ \ 0.363_{-0.006}^{+0.008}$%
\end{tabular}%
$ \\ \hline
$\lbrack 18,20.25]%
\begin{tabular}{l}
$SM$ \\ 
$Z^\prime $%
\end{tabular}%
$ & $%
\begin{tabular}{l}
$-0.735_{-0.009}^{+0.011}$ \\ 
$-0.494_{-0.002}^{+0.002}$%
\end{tabular}%
$ & $%
\begin{tabular}{l}
$ \ \ 0.226_{-0.042}^{+0.033}$ \\ 
$ \ \ 0.270_{-0.040}^{+0.043}$%
\end{tabular}%
$ & $%
\begin{tabular}{l}
$ \ \ 0.209_{-0.005}^{+0.004}$ \\ 
$ \ \ 0.371_{-0.004}^{+0.004}$%
\end{tabular}%
$ & $%
\begin{tabular}{l}
$ \ \ 0.791_{-0.004}^{+0.005}$ \\ 
$ \ \ 0.629_{-0.004}^{+0.004}$%
\end{tabular}%
$ & $%
\begin{tabular}{l}
$ \ \ 0.051_{-0.001}^{+0.001}$ \\ 
$ \ \ 0.079_{-0.001}^{+0.001}$%
\end{tabular}%
$ & $%
\begin{tabular}{l}
$ \ \ 0.686_{-0.002}^{+0.012}$ \\ 
$ \ \ 0.443_{-0.005}^{+0.007}$%
\end{tabular}%
$%
\end{tabular}%
\end{table}
\begin{table}
\caption{Average values of observables $\mathcal{P}_{i}$, $i= 2, . . . , 9$ (for real observables) for $\Lambda_b \to \Lambda (\to p\pi)\mu^{+}\mu^{-}$ in appropriate bins.}\label{pbin-values}
\begin{tabular}{lllllll}
\hline\hline
& $\ \ \ \ \ \ \left\langle \mathcal{P}_{2}\right\rangle $ & $\ \ \ \ \ \ \left\langle
\mathcal{P}_{3}\right\rangle $ & $\ \ \ \ \ \ \left\langle \mathcal{P}_{5}\right\rangle $ & $\ \
\ \ \ \ \left\langle \mathcal{P}_{6}\right\rangle $ & $\ \ \ \ \ \ \left\langle
\mathcal{P}_{8}\right\rangle $ & $\ \ \ \ \ \ \left\langle \mathcal{P}_{9}\right\rangle $ \\ 
\hline\hline
$\lbrack 0.1,2]%
\begin{tabular}{l}
$SM$ \\ 
$Z^\prime $%
\end{tabular}%
$ & \multicolumn{1}{||l}{$%
\begin{tabular}{l}
$ \ \ 0.240_{-0.176}^{+0.034}$ \\ 
$ \ \ 0.102_{-0.024}^{+0.004}$%
\end{tabular}%
$} & $%
\begin{tabular}{l}
$-0.076_{-0.002}^{+0.036}$ \\ 
$-0.056_{-0.002}^{+0.021}$%
\end{tabular}%
$ & $%
\begin{tabular}{l}
$-0.011_{-0.014}^{+0.054}$ \\ 
$ \ \ 0.019_{-0.002}^{+0.005}$%
\end{tabular}%
$ & $%
\begin{tabular}{l}
$-0.079_{-0.038}^{+0.178}$ \\ 
$-0.288_{-0.001}^{+0.005}$%
\end{tabular}%
$ & $%
\begin{tabular}{l}
$ \ \ 0.221_{-0.093}^{+0.003}$ \\ 
$ \ \ 0.107_{-0.065}^{+0.008}$%
\end{tabular}%
$ & $%
\begin{tabular}{l}
$-0.511_{-0.247}^{+0.056}$ \\ 
$-0.043_{-0.009}^{+0.001}$%
\end{tabular}%
$ \\ \hline
$\lbrack 1,2]%
\begin{tabular}{l}
$SM$ \\ 
$Z^\prime $%
\end{tabular}%
$ & \multicolumn{1}{||l}{$%
\begin{tabular}{l}
$ \ \ 0.458_{-0.166}^{+0.024}$ \\ 
$ \ \ 0.170_{0.030}^{0.004}$%
\end{tabular}%
$} & $%
\begin{tabular}{l}
$-0.090_{-0.002}^{+0.024}$ \\ 
$-0.079_{-0.001}^{+0.020}$%
\end{tabular}%
$ & $%
\begin{tabular}{l}
$ \ \ 0.001_{-0.014}^{+0.057}$ \\ 
$ \ \ 0.027_{-0.003}^{+0.008}$%
\end{tabular}%
$ & $%
\begin{tabular}{l}
$-0.044_{-0.029}^{+0.151}$ \\ 
$-0.191_{-0.102}^{+0.028}$%
\end{tabular}%
$ & $%
\begin{tabular}{l}
$ \ \ 0.257_{-0.048}^{+0.012}$ \\ 
$ \ \ 0.150_{-0.076}^{+0.007}$%
\end{tabular}%
$ & $%
\begin{tabular}{l}
$-0.231_{-0.222}^{+0.036}$ \\ 
$-0.046_{-0.003}^{+0.006}$%
\end{tabular}%
$ \\ \hline
$\lbrack 2,4]%
\begin{tabular}{l}
$SM$ \\ 
$Z^\prime $%
\end{tabular}%
$ & \multicolumn{1}{||l}{$%
\begin{tabular}{l}
$ \ \ 0.539_{-0.041}^{+0.001}$ \\ 
$ \ \ 0.202_{-0.003}^{+0.001}$%
\end{tabular}%
$} & $%
\begin{tabular}{l}
$-0.050_{-0.020}^{+0.011}$ \\ 
$-0.045_{-0.015}^{+0.007}$%
\end{tabular}%
$ & $%
\begin{tabular}{l}
$ \ \ 0.025_{-0.012}^{+0.050}$ \\ 
$ \ \ 0.033_{-0.004}^{+0.013}$%
\end{tabular}%
$ & $%
\begin{tabular}{l}
$ \ \ 0.042_{-0.019}^{+0.088}$ \\ 
$ \ \ 0.138_{-0.167}^{+0.040}$%
\end{tabular}%
$ & $%
\begin{tabular}{l}
$ \ \ 0.145_{-0.039}^{+0.084}$ \\ 
$ \ \ 0.089_{-0.006}^{+0.003}$%
\end{tabular}%
$ & $%
\begin{tabular}{l}
$-0.126_{-0.050}^{+0.000}$ \\ 
$-0.069_{-0.008}^{+0.028}$%
\end{tabular}%
$ \\ \hline
$\lbrack 4,6]%
\begin{tabular}{l}
$SM$ \\ 
$Z^\prime $%
\end{tabular}%
$ & \multicolumn{1}{||l}{$%
\begin{tabular}{l}
$ \ \ 0.516_{-0.013}^{+0.030}$ \\ 
$ \ \ 0.196_{-0.005}^{+0.013}$%
\end{tabular}%
$} & $%
\begin{tabular}{l}
$ \ \ 0.020_{-0.041}^{+0.015}$ \\ 
$ \ \ 0.024_{-0.031}^{+0.010}$%
\end{tabular}%
$ & $%
\begin{tabular}{l}
$ \ \ 0.049_{-0.010}^{+0.038}$ \\ 
$ \ \ 0.037_{-0.004}^{+0.013}$%
\end{tabular}%
$ & $%
\begin{tabular}{l}
$ \ \ 0.119_{-0.018}^{+0.064}$ \\ 
$ \ \ 0.427_{-0.115}^{+0.030}$%
\end{tabular}%
$ & $%
\begin{tabular}{l}
$-0.055_{-0.051}^{+0.149}$ \\ 
$-0.034_{-0.021}^{+0.053}$%
\end{tabular}%
$ & $%
\begin{tabular}{l}
$-0.155_{-0.020}^{+0.051}$ \\ 
$-0.100_{-0.011}^{+0.036}$%
\end{tabular}%
$ \\ \hline
$\lbrack 6,8]%
\begin{tabular}{l}
$SM$ \\ 
$Z^\prime $%
\end{tabular}%
$ & \multicolumn{1}{||l}{$%
\begin{tabular}{l}
$ \ \ 0.463_{-0.016}^{+0.041}$ \\ 
$ \ \ 0.176_{-0.005}^{+0.014}$%
\end{tabular}%
$} & $%
\begin{tabular}{l}
$ \ \ 0.082_{-0.035}^{+0.014}$ \\ 
$ \ \ 0.089_{-0.025}^{+0.009}$%
\end{tabular}%
$ & $%
\begin{tabular}{l}
$ \ \ 0.067_{-0.009}^{+0.027}$ \\ 
$ \ \ 0.040_{-0.003}^{+0.010}$%
\end{tabular}%
$ & $%
\begin{tabular}{l}
$ \ \ 0.166_{-0.020}^{+0.057}$ \\ 
$ \ \ 0.579_{-0.053}^{+0.016}$%
\end{tabular}%
$ & $%
\begin{tabular}{l}
$-0.231_{-0.049}^{+0.133}$ \\ 
$-0.143_{-0.024}^{+0.060}$%
\end{tabular}%
$ & $%
\begin{tabular}{l}
$-0.220_{-0.025}^{+0.066}$ \\ 
$-0.125_{-0.011}^{+0.032}$%
\end{tabular}%
$ \\ \hline
$\lbrack 14,16]%
\begin{tabular}{l}
$SM$ \\ 
$Z^\prime $%
\end{tabular}%
$ & \multicolumn{1}{||l}{$%
\begin{tabular}{l}
$ \ \ 0.318_{-0.005}^{+0.009}$ \\ 
$ \ \ 0.136_{-0.002}^{+0.002}$%
\end{tabular}%
$} & $%
\begin{tabular}{l}
$ \ \ 0.214_{-0.006}^{+0.003}$ \\ 
$ \ \ 0.234_{-0.004}^{+0.003}$%
\end{tabular}%
$ & $%
\begin{tabular}{l}
$ \ \ 0.135_{-0.001}^{+0.002}$ \\ 
$ \ \ 0.066_{-0.000}^{+0.001}$%
\end{tabular}%
$ & $%
\begin{tabular}{l}
$ \ \ 0.158_{-0.008}^{+0.008}$ \\ 
$ \ \ 0.672_{-0.001}^{+0.000}$%
\end{tabular}%
$ & $%
\begin{tabular}{l}
$-0.564_{-0.012}^{+0.022}$ \\ 
$-0.368_{-0.009}^{+0.013}$%
\end{tabular}%
$ & $%
\begin{tabular}{l}
$-0.365_{-0.007}^{+0.014}$ \\ 
$-0.169_{-0.004}^{+0.007}$%
\end{tabular}%
$ \\ \hline
$\lbrack 16,18]%
\begin{tabular}{l}
$SM$ \\ 
$Z^\prime $%
\end{tabular}%
$ & \multicolumn{1}{||l}{$%
\begin{tabular}{l}
$ \ \ 0.317_{-0.002}^{+0.005}$ \\ 
$ \ \ 0.146_{-0.001}^{+0.002}$%
\end{tabular}%
$} & $%
\begin{tabular}{l}
$ \ \ 0.202_{-0.003}^{+0.002}$ \\ 
$ \ \ 0.240_{-0.003}^{+0.002}$%
\end{tabular}%
$ & $%
\begin{tabular}{l}
$ \ \ 0.163_{-0.000}^{+0.000}$ \\ 
$ \ \ 0.085_{-0.000}^{+0.000}$%
\end{tabular}%
$ & $%
\begin{tabular}{l}
$ \ \ 0.090_{-0.001}^{+0.002}$ \\ 
$ \ \ 0.641_{-0.003}^{+0.002}$%
\end{tabular}%
$ & $%
\begin{tabular}{l}
$-0.527_{-0.005}^{+0.013}$ \\ 
$-0.371_{-0.004}^{+0.006}$%
\end{tabular}%
$ & $%
\begin{tabular}{l}
$-0.338_{-0.003}^{+0.008}$ \\ 
$-0.165_{-0.002}^{+0.003}$%
\end{tabular}%
$ \\ \hline
$\lbrack 18,20.25]%
\begin{tabular}{l}
$SM$ \\ 
$Z^\prime $%
\end{tabular}%
$ & \multicolumn{1}{||l}{$%
\begin{tabular}{l}
$ \ \ 0.313_{-0.002}^{+0.001}$ \\ 
$ \ \ 0.187_{-0.002}^{+0.002}$%
\end{tabular}%
$} & $%
\begin{tabular}{l}
$ \ \ 0.135_{-0.005}^{+0.000}$ \\ 
$ \ \ 0.210_{-0.003}^{+0.002}$%
\end{tabular}%
$ & $%
\begin{tabular}{l}
$ \ \ 0.180_{-0.003}^{+0.002}$ \\ 
$ \ \ 0.127_{-0.000}^{+0.000}$%
\end{tabular}%
$ & $%
\begin{tabular}{l}
$-0.032_{-0.007}^{+0.001}$ \\ 
$ \ \ 0.531_{-0.008}^{+0.007}$%
\end{tabular}%
$ & $%
\begin{tabular}{l}
$-0.350_{-0.002}^{+0.013}$ \\ 
$-0.323_{-0.002}^{+0.002}$%
\end{tabular}%
$ & $%
\begin{tabular}{l}
$-0.222_{-0.002}^{+0.009}$ \\ 
$-0.143_{-0.001}^{+0.001}$%
\end{tabular}%
$%
\end{tabular}%
\end{table}%

\section{Conclusion}
In this study we have investigated the full four-folded angular distributions for the semileptonic $b$-baryon decay $ \Lambda_b \rightarrow \Lambda(\rightarrow p \pi)\mu^{+}\mu^{-}$ in the SM and $Z^{\prime}$ model. At the quark level, this decay is mediated by the quark level transition $b \to s \mu^{+}\mu^{-}$ that is same for the well studied meson decay $B \to K^{*}\mu^{+}\mu^{-}$ decay. For $\Lambda_b \to \Lambda$ transitions, we have used the high precision form factors calculated in the lattice QCD using $2 +1$ dynamical flavors along with the factorizable non-local matrix elements 
of the four quark operators $\mathcal{O}_{1-6}$ and $\mathcal{O}_{8}^g$ encoded into effective Wilson coefficients $C_7^{eff}(s)$ and $C_9^{eff}(s)$. By using them we have calculated numerically the differential branching ratio $\frac{d\mathcal{B}}{ds}$, the lepton-, hadron-, combined hardron-lepton forward-backward asymmetries $(A^{\ell}_{FB}, A^{\Lambda}_{FB}, A^{\ell\Lambda}_{FB})$, the various asymmetry parameters $(\alpha)$'s, the fractions of longitudinal $(F_{L})$, transverse $(F_{T})$ polarized dimuons and different angular asymmetry observables denoted by $\mathcal{P}$ in different bins of $s$.
\begin{itemize}
\item In the low-recoil bin $s\in [15, 20]$ GeV$^2$ the form factors from lattice are known to be most precisely, the results of $\frac{d\mathcal{B}}{ds}$ in $Z^{\prime}$ model lies close to the experimental measurements in this bin. The SM results are significantly smaller than the measurements in this low-recoil bin.
\item In large-recoil region the results of hadron-side forward-backward asymmetry $(A^{\Lambda}_{FB})$ is significantly away from the experimental observations both for the SM and $Z^{\prime}$ model. However, in the low-recoil region the results of the SM lies close to the experimental observations.
\item The experimental measurements of lepton-side forward-backward asymmetry $(A^{\ell}_{FB})$ both in the low- and large-recoil regions have significantly large errors. However, in the bin $s\in [15, 20]$ GeV$^2$ the lower limit is comparable to the $Z^{\prime}$ model. We hope in future, when the statistics of the data will be improved it will help us to find the signatures of extra neutral $Z^{\prime}$ boson.
\item We have also predicted the values of lepton-hardon combined forward-backward asymmetry $(A^{\ell\Lambda}_{FB})$ both in the SM and $Z^{\prime}$ model. It has been found that in the low-recoil bin the value of $Z^{\prime}$ model deviates significantly from the SM result.
\item The longitudinal polarization fraction $F_{L}$ of the dimuon system is measured experimentally where the statistics is not good enough in the large-recoil bin as compared to the low-recoil region. In the region $s\in [1, 6]$ GeV$^2$ the central value of the SM is compatible with the central value of the experimental measurements. However, in the bin $s\in [15, 20]$ GeV$^2$, where uncertainties in the form factors are better controlled, the experimental observations favor the results of the $Z^{\prime}$ model.
\item In line with these asymmetries, we have also calculated the transverse polarization fraction of dimuon system $F_{T}$, the asymmetry parameters $\alpha$'s and different angular asymmetry observables $\mathcal{P}_{i}$ for $i = 1, \cdots , 9$ in the SM and $Z^{\prime}$ model. We have found significantly large values of some of these observables that can be measured in the future at LHCb and Belle II.
\end{itemize}
In the end we would like to emphasis that some of the asymmetries calculated here were also reported in the SM and alligned 2HDM in ref. \cite{Hu:2017qxj} and our SM results matches with these results. We hope that in future, the precise measurement of some of the asymmetries reported here in the four-folded distribution of $ \Lambda_b \rightarrow \Lambda(\rightarrow p \pi)\mu^{+}\mu^{-}$ decay, in fine bins of the $s$ at the LHCb and Belle II will help us to test the SM predictions in $\Lambda_b$ decays with significantly improved statistics. It will also give us a chance to hunt for the indirect signals of NP arising due to the neutral $Z^{\prime}$ boson especially where SM is at mismatch with the experimental predictions.
\section*{acknowledgements}
The authors would like to thank Wei Wang and Ishtiaq Ahmed for the useful discussions and suggestions to improve the work. In addition, MJA and SS would like to thank D. Van Dyk for his help to understand the uncertainties arising due to different inputs. MJA would like to acknowledge the support by Quaid-i-Azam University through the University Research
Funds and the funds from the Centre for Future High Energy Physics, Beijing, China through its visiting scientist scheme.

\newpage
\section*{Appendix}
In the rest frame of the decaying $\Lambda_b$ baryon, the momentum of daughter baryon $\Lambda$ is defined as
\begin{equation*}
p_2= (m_{\Lambda_b}-q_0, 0, 0, |\bf{q}|)
\end{equation*}
where $m_{\Lambda_b}$ is the mass of the $\Lambda_b$ baryon. The lepton polarization vectors in the dilepton rest frame are given as
\begin{eqnarray}
&\epsilon^{\mu}_+= \frac{1}{\sqrt{2}} (0, 1, -i, 0 ) ,\quad\quad\quad\quad
\epsilon^{\mu}_-= \frac{1}{\sqrt{2}} (0, -1, -i, 0 )& \notag \\
&\epsilon^{\mu}_t=  (1, 0, 0, 0) ,\quad\quad\quad\quad\quad\quad
\epsilon^{\mu}_0= (0, 0, 0, 1) \ \ \ \ \ \ \ \ \notag &
\end{eqnarray}
and the corresponding 
Lepton momentum vectors are \cite{38}
\begin{eqnarray*}
q^{\mu}_1 &=& (E_{\ell}, -|\bf{q}|\sin{\theta_l}, 0, -|\bf{q}|\cos{\theta_l})\\
q^{\mu}_2 &=& (E_{\ell}, |\bf{q}|\sin{\theta_l}, 0, |\bf{q}|\cos{\theta_l})
\end{eqnarray*}
with$E_{\ell}=\frac{\sqrt{s}}{2}$ and
$|{\bf q}|= E^2_{\ell}-m^2_{\ell}$. \\

The helicity amplitudes for the decay $\Lambda_b \rightarrow \Lambda $ transitions can be expressed in terms of the form factors as \cite{38}
\begin{eqnarray*}
H_{t\left( +1/2,+1/2\right) }^{V} &=&H_{t\left( -1/2,-1/2\right)
}^{V}=f_{0}\left( s\right) \frac{m_{\Lambda _{b}}-m_{\Lambda }}{\sqrt{s}}%
\sqrt{s_{+}},  \notag \\
H_{0\left( +1/2,+1/2\right) }^{V} &=&H_{0\left( -1/2,-1/2\right)
}^{V}=f_{+}\left( s\right) \frac{m_{\Lambda _{b}}+m_{\Lambda }}{\sqrt{s}}%
\sqrt{s_{-}},  \notag \\
H_{+\left( -1/2,+1/2\right) }^{V} &=&H_{-\left( +1/2,-1/2\right)
}^{V}=-f_{\bot }\left( s\right) \sqrt{2s_{-}},  \\
H_{t\left( +1/2,+1/2\right) }^{A} &=&-H_{t\left( -1/2,-1/2\right)
}^{A}=g_{0}\left( s\right) \frac{m_{\Lambda _{b}}+m_{\Lambda }}{\sqrt{s}}%
\sqrt{s_{-}},  \notag \\
H_{0\left( +1/2,+1/2\right) }^{A} &=&-H_{0\left( -1/2,-1/2\right)
}^{A}=g_{+}\left( s\right) \frac{m_{\Lambda _{b}}-m_{\Lambda }}{\sqrt{s}}%
\sqrt{s_{+}},  \notag \\
H_{+\left( -1/2,+1/2\right) }^{A} &=&-H_{-\left( +1/2,-1/2\right)
}^{A}=-g_{\bot }\left( s\right) \sqrt{2s_{+}} \\
H_{0\left( +1/2,+1/2\right) }^{T} &=&H_{0\left( -1/2,-1/2\right)
}^{T}=-h_{+}\left( s\right) \sqrt{s} \sqrt{s_{-}},  \notag \\
H_{+\left( -1/2,+1/2\right) }^{T} &=&H_{-\left( +1/2,-1/2\right)
}^{T}=h_{\bot }\left( s\right) \left( m_{\Lambda _{b}}+m_{\Lambda }\right) 
\sqrt{2s_{-}},  \\
H_{0\left( +1/2,+1/2\right) }^{T5} &=&-H_{0\left( -1/2,-1/2\right) }^{T5}=%
\widetilde{h}_{+}\left( s\right) \sqrt{s}\sqrt{s_{+}},  \notag \\
H_{+\left( -1/2,+1/2\right) }^{T5} &=&-H_{-\left( +1/2,-1/2\right) }^{T5}=-%
\widetilde{h}_{\bot }\left( s\right) \left( m_{\Lambda _{b}}-m_{\Lambda
}\right) \sqrt{2s_{+}},  
\end{eqnarray*}%
where $f_{0}, f_+$ and $f_{\perp}$ denotes time-like, longitudinal and transverse components of vector currents. The kinematic functions used in the above equation are defined as
$s_{\pm} \equiv (m_{\Lambda_b} \pm m_{\Lambda})^2 - s$.

The transversity amplitude can be written in terms of helicity amplitudes as \cite{38}
\begin{eqnarray}
A_{\bot _{1}}^{L\left( R\right) } &=&+\sqrt{2}\mathcal{N}\left( \left( C_{9}^{+}\mp
C_{10}^{+}\right) H_{+\left( -1/2,+1/2\right) }^{V}-\frac{2m_{b}
	C_{7}^+ }{s}H_{+\left( -1/2,+1/2\right) }^{T}\right) , 
\notag \\
A_{\Vert _{1}}^{L\left( R\right) } &=&-\sqrt{2}\mathcal{N}\left( \left( C_{9}^{-}\mp
C_{10}^{-}\right) H_{+\left( -1/2,+1/2\right) }^{A}+\frac{2m_{b}
	C_{7}^- }{s}H_{+\left( -1/2,+1/2\right) }^{T5}\right) , \\
A_{\bot _{0}}^{L\left( R\right) } &=&+\sqrt{2}\mathcal{N}\left( \left( C_{9}^{+}\mp
C_{10}^{+}\right) H_{0\left( +1/2,+1/2\right) }^{V}-\frac{2m_{b}
	C_{7}^+ }{s}H_{0\left( +1/2,+1/2\right) }^{T}\right) , 
\notag \\
A_{\Vert _{0}}^{L\left( R\right) } &=&-\sqrt{2}\mathcal{N}\left( \left( C_{9}^{-}\mp
C_{10}^{-}\right) H_{0\left( +1/2,+1/2\right) }^{A}+\frac{2m_{b}
	C_{7}^- }{s}H_{0\left( +1/2,+1/2\right) }^{T5}\right) .
\notag
\end{eqnarray}
where $ \mathcal{N}=G_{F}V_{tb}V_{ts}^{\ast }\alpha _{e}\sqrt{\frac{s \lambda^{1/2} \left(
			m_{\Lambda _{b}}^{2},\; m_{\Lambda }^{2},\; s\right)}{3\cdot 2^{11}m_{\Lambda
			_{b}}^{3}\pi ^{5}}} $ and
\begin{eqnarray}
C_{9}^{+}=C_{9}+C_{9}^{\prime },\quad C_{9}^{-}=C_{9}-C_{9}^{\prime
},\quad C_{10}^{+}=C_{10}+C_{10}^{\prime }, \quad C_{10}^{-}=C_{10}-C_{10}^{\prime } ,\quad C^{+}_7= C_7+C_7^{\prime},\quad C^{-}_7= C_7-C_7^{\prime}. \notag
\end{eqnarray}%
In case of SM and $Z^{\prime}$ model, all the primed Wilson coefficients are zero.

The angular variable $K_{l,m}$ with $l$ and $m$ denoting the relative angular momentum and its third component for $p \pi$ and $ \mu^+ \mu^-$ systems, respectively that are introduced in Eq. (\ref{6}) can be written in terms of transversality amplitudes as \cite{38}:
\begin{eqnarray}
K_{1ss}\left( s\right)  &=&\frac{1}{4}\left[ \left\vert A_{\bot
_{1}}^{R}\right\vert ^{2}+\left\vert A_{\Vert _{1}}^{R}\right\vert
^{2}+2\left\vert A_{\bot _{0}}^{R}\right\vert ^{2}+2\left\vert A_{\Vert
_{0}}^{R}\right\vert ^{2}+\left( R\leftrightarrow L\right) \right] , K_{1cc}\left( s\right)  = \frac{1}{2}\left[ \left\vert A_{\bot
_{1}}^{R}\right\vert ^{2}+\left\vert A_{\Vert _{1}}^{R}\right\vert
^{2}+\left( R\leftrightarrow L\right) \right] , \notag  \\
K_{1c}\left( s\right)  &=&-{Re}\left\{ A_{\bot _{1}}^{R}A_{\Vert
_{1}}^{\ast R}-\left( R\leftrightarrow L\right) \right\},  K_{2ss}\left( s\right)  = \frac{\alpha }{2} {Re}\left\{ A_{\bot
_{1}}^{R}A_{\Vert _{1}}^{\ast R}+2A_{\bot _{0}}^{R}A_{\Vert _{0}}^{\ast
R}+\left( R\leftrightarrow L\right) \right\} , \notag \\
K_{2cc}\left( s\right)  &=&+\alpha {Re}\left\{ A_{\bot
_{1}}^{R}A_{\Vert _{1}}^{\ast R}+\left( R\leftrightarrow L\right) \right\}, K_{2c}\left( s\right)  = -\frac{\alpha }{2}\left[ \left\vert A_{\bot
_{1}}^{R}\right\vert ^{2}+\left\vert A_{\Vert _{1}}^{R}\right\vert
^{2}-\left( R\leftrightarrow L\right) \right] \notag , \\ 
K_{2sc}\left( s\right)  &=&+\frac{\alpha }{\sqrt{2}} {Im}\left\{ A_{\bot
_{1}}^{R}A_{\bot _{0}}^{\ast R}-A_{\Vert _{1}}^{R}A_{\Vert _{0}}^{\ast
R}+\left( R\leftrightarrow L\right) \right\} , K_{3s}\left( s\right)  = \frac{\alpha }{\sqrt{2}} {Im}\left\{ A_{\bot
_{1}}^{R}A_{\Vert _{0}}^{\ast R}-A_{\Vert _{1}}^{R}A_{\bot _{0}}^{\ast
R}-\left( R\leftrightarrow L\right) \right\} \notag , \\
K_{4sc}\left( s\right)  &=&+\frac{\alpha }{\sqrt{2}} {Re}\left\{ A_{\bot
_{1}}^{R}A_{\Vert _{0}}^{\ast R}-A_{\Vert _{1}}^{R}A_{\bot _{0}}^{\ast
R}+\left( R\leftrightarrow L\right) \right\} ,  K_{4s}\left( s\right)  = \frac{\alpha }{\sqrt{2}} {Re}\left\{ A_{\bot
_{1}}^{R}A_{\bot _{0}}^{\ast R}-A_{\Vert _{1}}^{R}A_{\Vert _{0}}^{\ast
R}-\left( R\leftrightarrow L\right) \right\} .\notag
\end{eqnarray}%

 \end{document}